\documentclass{svjour3}                     
\smartqed  
\usepackage{graphicx}
\usepackage[english]{babel}
\usepackage{url}
\usepackage{hyperref}
\usepackage{verbatim}
\usepackage{hyperref}
\usepackage{placeins}

\journalname{EPJ Data Science}
\begin{document}

\title{Complex delay dynamics on railway networks
}

\subtitle{from universal laws to realistic modelling}

\titlerunning{Complex delay dynamics on railway networks}        

\author{Bernardo Monechi      \and
        Pietro Gravino \and
        Riccardo Di Clemente \and
        Vito D.~P.~Servedio
}


\institute{B.~Monechi \at
              Institute for Scientific Interchange Foundation, 
              Via Alassio 11/c, 10126, Turin, Italy \\ and SONY Computer Science Lab, 6 Rue Amyot, Paris, France, 75005, France
              \email{mone.berna@gmail.com}           
       \and
           P.~Gravino \at
              "Sapienza" University of Rome, 
              Piazzale Aldo Moro 5, 00185, Roma\\ and SONY Computer Science Lab, 6 Rue Amyot, Paris, 75005, France
              \email{pietro.gravino@gmail.com}   
       \and
           R.~Di Clemente \at
              University College London, 
              The Bartlett Centre for Advanced Spatial Analysis, 
              London, WC1E 6BT, United Kingdom
              \at
              Massachusetts Institute of Technology, 
              77 Massachusetts Ave, MA 02139 - Cambridge, USA\\
              \email{r.diclemente@ucl.ac.uk}   
       \and
           V.~D.~P.~Servedio \at
              Complexity Science Hub Vienna, 
              Josefst\"adter Str. 39, 1080 Vienna, Austria\\
              {\email{servedio@csh.ac.at}}   
}

\date{Received: date / Accepted: date}

\maketitle

\begin{abstract}

\noindent 
Railways are a key infrastructure for any modern country. 
The reliability and resilience of this peculiar transportation system may be challenged by different shocks such as disruptions, strikes and adverse weather conditions.
These events compromise the correct functioning of the system and trigger the spreading of delays into the railway network on a daily basis. 
Despite their importance, a general theoretical understanding of the underlying causes of these disruptions is still lacking. 
In this work, we analyse the Italian and German railway networks by leveraging on the train schedules and actual delay data retrieved during the year 2015. 
We use {these} data to infer simple statistical laws ruling the emergence of localized delays in different areas of the network and we model the spreading of these delays throughout the network by exploiting a framework inspired by epidemic spreading models. 
Our model offers a fast and easy tool for the preliminary assessment of the {effectiveness of} traffic handling policies, and of the railway {network} criticalities.

\end{abstract}

\section{Introduction}
\label{sec:intro}

Transportation networks are a critical infrastructure with an enormous impact on local, national, and international economies \cite{banister1997sustainable,limao2001infrastructure}.
At {the} micro-level, the commuting time {impacts} the economy and the shape {of cities} \cite{newman1999sustainability}, directly  influencing our life style and choices \cite{train2009discrete}. 
At {the} macro-level the travel time regulates trade and stimulates economic activities \cite{nelson2008travel}.
Railroads, when correctly developed, foster interregional trades within the country, and increase income levels within the city boundaries \cite{donaldson2018railroads}. 
Whether we are traveling in crowded {coaches} of the commuters rail {during} rush {hours} or in the long-medium high speed train trips, the reliability of the transportation system is a key factor in determining travel behavior \cite{carrion2012value}. 
The travel time {unreliability} in rail service can have substantial {consequences for its users} \cite{noland2002travel} and the growth of cities \cite{banister2008sustainable}.
{Delays are}  one of the causes that {corrupt}  the travel time reliability of the transportation systems.
{To address delay propagation,} traditional approaches apply {either} stochastic models  \cite{meester2007stochastic} or use propagation {algorithms} on a timed event graph representation of a scheduled railway system  \cite{goverde2010delay}. 
Complex equations describing delay were obtained and solved by means of iterative refinement algorithms to predict positive delays in urban trains \cite{higgins1998modeling}, while traffic control models were proposed to manage the safety of timetables after perturbations occur \cite{d2007conflict}.

The development of models {addressing} the dynamics of trains over railway networks are usually focused either on delimited aspects, e.g.{,} single lane dynamics \cite{li2005cellular} or the issuing of schedule  \cite{cacchiani2010scheduling,csahin1999railway}, or are comprehensively including a wide variety of different microscopic ingredients that can be hardly validated by real data \cite{giua2008modeling,middelkoop2001simone}. 
In the last years a new perspective bloomed to provide a different understanding of delay dynamics through the lens of Complex Systems.
Railway networks have largely been studied and exhibit small-world scaling properties  \cite{sen2003small}, their topology have geo-spatial restriction  \cite{erath2009graph}  and  structural redundancy  \cite{bhatia2015network}. 
On the other end, a railway delay dynamics framework within a complex systems approach is still lacking. 
Our aim is to develop a model capable of reproducing  the actual delay structure and the emergence of congested areas, by exploiting some essential ingredients without relying on the detailed knowledge of the microscopic mechanisms underlying delay generation. 
A similar approach has already been successfully applied in the context of the Air Transport System of USA and Europe \cite{fleurquin2013systemic,campanelli2014modeling}, where interaction models were used to characterize and forecast the spreading of delays among flights.
These modelling schemes are driven by the necessity of simple and novel frameworks to study transportation systems, allowing  fast scenario simulations for schedule testing, which could be easily applied to resilience studies that are nowadays performed independently of the dynamics taking place over the network \cite{bhatia2015network}. 
\newline 
Following a similar approach, we develop a new data driven model of delay propagation among trains diffusing over a railway network. 
We start by inferring universal laws governing the emergence of delays over the railway network as a consequence of the occurrence of adverse conditions that are not depending on the (possible) interactions between trains. 
Hence, inspired by models for epidemic spreading \cite{pastor2001epidemic}, we introduce the \emph{Delay Propagation  Model} as a novel framework to asses and  test how such emerging delays spawn and spread over the network. 
We name the first kind of delays as ``exogenous delays'', while the second ones stemming from the interaction between trains as ``endogenous delays''. 
While the occurrence of exogenous delays can be detected and analysed from our data,
the endogenous delays, on the contrary, since the (supposed) direct interaction between trains cannot be inferred from the dataset we collected, will be modeled with a one parameter interaction.
The effects and consistency of such endogenous delay modelling scheme will be checked \emph{a posteriori} by means of numerical simulations. 
Our model is close to an SIS model of epidemic spreading \cite{pastor2001epidemic}, since trains can either get infected by delay, recover from it and then get infected again.
We show that this model is capable of reproducing the empirical distribution of delays measured in the data as well as  the emergence of large congested areas. 
Moreover, we show that the removal of the delay propagation mechanism prevents the modeled system from generating large disruptions, hence strongly suggesting the existence of this kind of interaction in the real system.
Finally, we propose an application of our model in studying a scenario where the propagation of a localized delay leads to the  emergence of a vast non-functioning area in the Italian Railway Network.\\
The paper is structured as follows: in the first section we describe the dataset and some empirical findings that inspired the assumptions used in the development of the model; 
in the second section we discuss the model in detail and will show its capability of reproducing empirical findings such as the distribution of delays and the size of congested areas; we will also demonstrate the usefulness  of the model as a scenario simulation tool; in the last section we will summarize our findings and discuss possible future developments.
\section{Methods}
\subsection{Data}
\begin{figure}[htbp]
	\begin{center}
		\includegraphics[width=.9\textwidth]{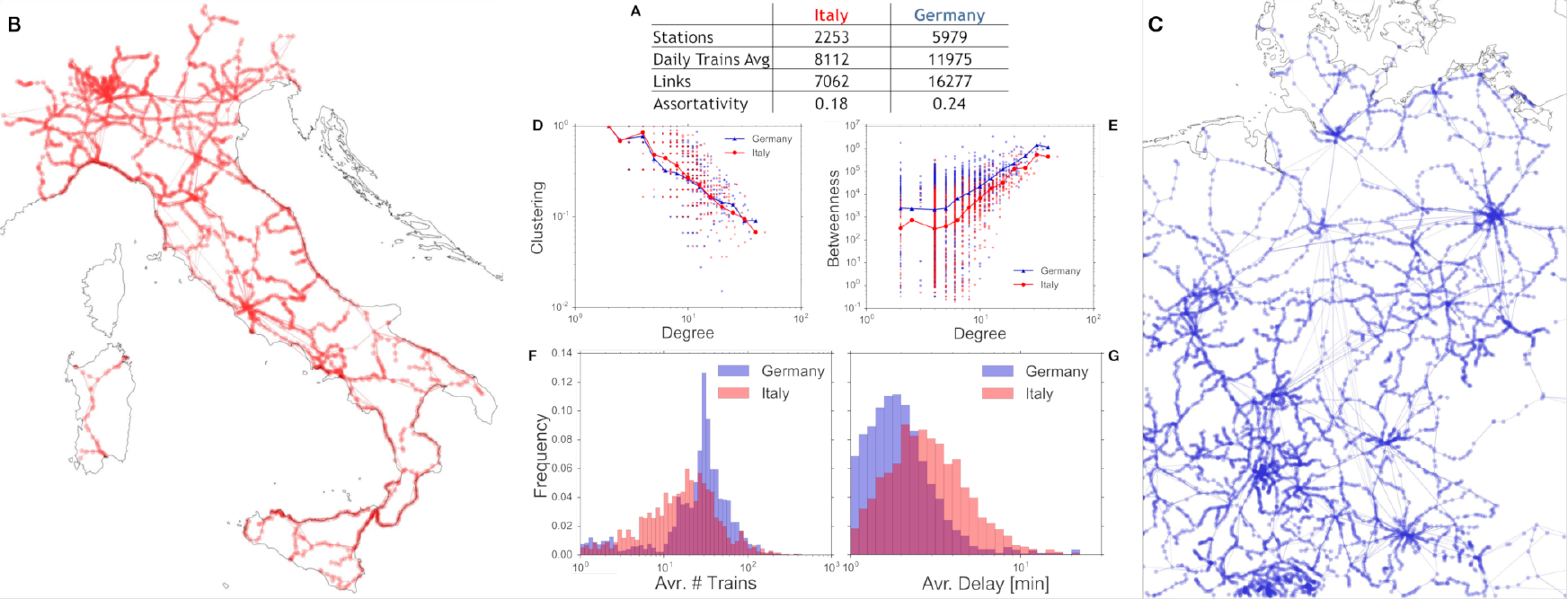}
		\caption{
			Italian and German Railway Networks and Traffic. 
			(A) A simple comparison of datasets and basic networks properties. 
			(B,C) Railway Networks shown on a map. Nodes of the network represent stations, which are linked by a straight line whenever there is a train going from one to another without intermediate stops. 
			(D) A scatter plot of the local clustering coefficient against the degree of nodes. Each dot represents a station and is collocated depending on its properties. 
			Lines represent the averaged clustering as a function of the degree of the nodes in the two networks. 
			(E) Similar to (D), but with betweenness centrality in place of clustering.
			(F) Comparison of the Italian and German distributions of the average number of trains per day travelling through a station.
			(G) Comparison between the Italian and German distributions of the average delay at stations. 
			All the analyses are made over the whole 2015. The Italian distribution has been made transparent in order to see the overlap with the German one.
		}
		\label{fig:networks}
	\end{center}
\end{figure}

In order to extract relevant patterns related with the emergence of train delays, we collected and analysed data about the daily operations of the Italian and German Railway Systems during the year 2015. 
Such data were collected in the first case through the ViaggiaTreno website \cite{viaggiatreno} and for the second one through the OpenDataCity website \cite{opendatacity}. 
The information we acquired allowed for a complete reconstruction of the schedules of the trains, the structure of the Railway Network and the delays that affected the trains during their movements. For more information about the data collection please refer to Supplementary Information (SI).
The choice of the Italian and German Systems as subjects for our studies lies in their similarities in structure and management. 
In fact, their networks have remarkable and comparable sizes (41,315 and 16,723~km, respectively) and densities ($8.22$ and $12.46$ km$^2$ per km of tracks, respectively \cite{cia-world-factbook}). 
Moreover, these two countries share also a crucial characteristic: in both railway systems traffic is handled mainly by a single national company. In other countries with similar networks, like in the United Kingdom, the railway network is managed by different companies resulting in a more complex system where trains are additionally subject to the commercial policies of different operators.
Fig.~\ref{fig:networks} shows a set of topological analyses of the two  railway networks, together with a preliminary analysis of the traffic load and delays in the systems. 
Following existing literature, we refer to a railway network as the network whose nodes represent stations that are connected by a link whenever there is a train connecting them with two consecutive stops in its schedule. 
The network is directed since it is possible that a connection between two stations is travelled just in one direction. 
In this paper we refer to the nodes in a railway network either as ``nodes'' or as ``stations''. 
Moreover, the action performed by a train travelling from station A to station B will be referred as ``travelling over the link'' connecting A and B.
In both networks the distribution of the degree $k$ is peaked on $k=4$ and for larger degrees has exponential-like decrease   \cite{sen2003small,kurant2006extraction}  proportional to $e^{-k/k_0}$ with $k_0 \simeq 4.5 \pm 0.1$ . 
This finding is in agreement with other geographical networks \cite{barthelemy2011spatial}.
The network assortativity (Fig.~\ref{fig:networks}A)  \cite{newman2002assortative} is for Italy and Germany, $0.18$ and $0.24$ respectively. 
This indicates that while there is a slight preference for stations with the same degree to be connected each other, yet the various degrees are mostly mixing, i.e., smaller and larger stations are typically connected.
The local clustering coefficient (Fig.~\ref{fig:networks}D),  defined as the fraction of pairs of neighbours of a given station that are connected over all pairs of neighbours of that station \cite{newman2003structure}, can be used to infer the redundancy of the network. When a disruption occurs, a station with an high clustering coefficient can be bypassed easily. 
Complementarily, betweenness centrality (Fig.~\ref{fig:networks}E) is a measure of the centrality of a node in a network based on the number of shortest paths that pass through it \cite{newman2003structure}. This measure highlights how strategic a given station is at a global level.

The clustering and the betweenness outline the typical small word topology of transportation networks \cite{sen2003small}.
German stations  show a distribution between $10$ and $100$ trains per day, narrower that the train distribution of the  
Italian stations, which instead display a broader distribution, suggesting a more heterogeneous handling of train traffic load (Fig.~\ref{fig:networks}F). 
Finally, Fig.~\ref{fig:networks}G shows the histogram of the average delays of trains aggregated by stations for the two nations. Both distributions show a peak and  heavy tail, whereas compared to the German distribution, the Italian distribution is clearly shifted towards higher delays. 
The topological similarities between the two networks suggest that the differences in delays and traffic load might be the result of differences in the trains dynamics.
\subsection{Train interaction on railways networks}
\begin{figure}[t]
	\begin{center}
		\includegraphics[width=\textwidth]{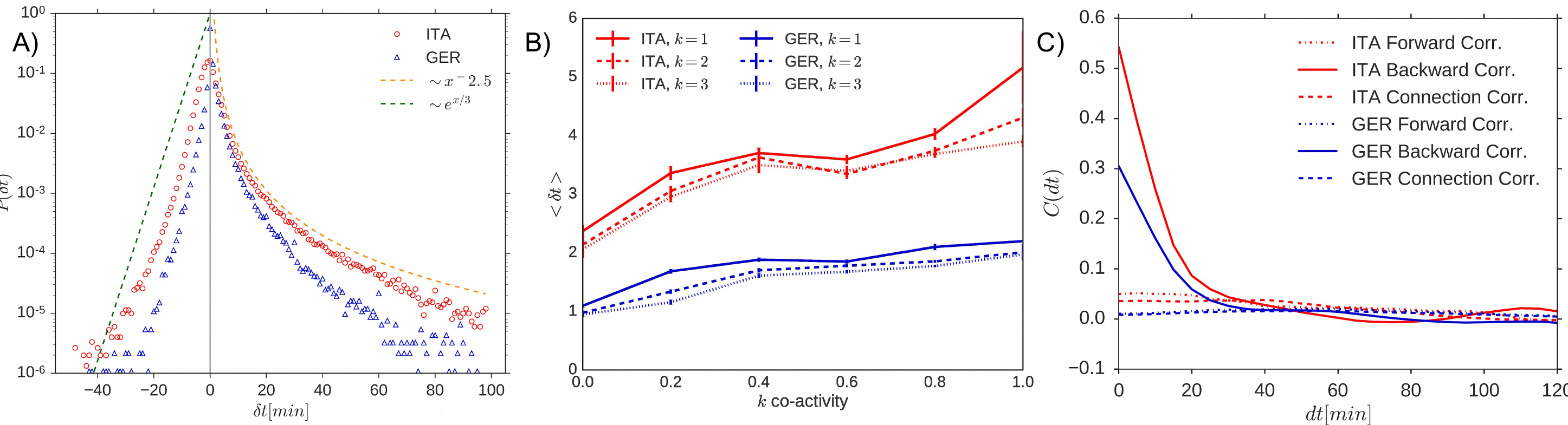}
		\caption{
			Delays and train interactions. 
			(A) Distributions of the delay at the final and at the intermediate stations trains in both Italian and German datasets. Departure delays at the initial station has been subtracted, hence these delays originated during the travel. As a guide for the eye, a power-law has been added in the positive delay area and an exponential in the advance area (negative delay). 
			(B) Average delay of a link as a function of the link $k$-order co-activities, with Spearman's correlation coefficients showing the existence of a relation between the co-activity and the average delay. The curves have been obtained by binning the data on the x-axis at steps of $0.2$ and reporting the average and the standard error.
			(C) Cross-Correlations $C(dt)$ from equation (\ref{eq:crosscorr}) as a function of $dt$ for the average delay time series of pairs of nearby links in the railway networks in the ``Forward'' and ``Backward'' configurations (shown in Fig.~\ref{fig:propagation}), and in the ``Connection'' configuration .
		}
		\label{fig:delays}
	\end{center}
\end{figure}

To focus on the analysis of delays, their outbreak and evolution, we choose  trains instead of stations as our reference systems.
The intermediate delays for a train $i$ travelling from station $A$ to station $B$ on the link $e$ is defined as $\Delta_i t(e)$, 
The departure delays at the initial station have been subtracted to analyse only the delays that have been generated during the travel. 
Hence $\Delta_i t(e)$ can be negative  whenever the train is in advance, resulting in the train waiting at the station for the correct time of departure. 
Fig.~\ref{fig:delays}A shows the delay distributions for both national systems, considering the delays at intermediate stations along the path of a train, or just at the final station. 
We observe similar shapes of these distributions, with the Italian one exhibiting broader tails than the German. 
More than $10\%$ of the train stops are on-time. 
The distributions of both countries exhibit an asymmetric pattern: the right tail (labeled Delay) shows a power-law like behaviour compatible with a q-exponential distribution \cite{Briggs2007modelling}, while the left tail (labelled Advance) has an exponential steeper slope. 
We expect this distribution to be the result of the interplay between the occurrence of adverse conditions and the interaction between trains, influencing their dynamics. 
Despite the fact that the microscopic details in our data do not allow for a precise investigation of possible interactions between trains, we can highlight how the possibility of interaction might affect the delays in railway networks.
Hence, we study the relation between  the \emph{first order co-activity} of a link, i.e., the probability that at a certain time $t$ (with time steps of $30$ minutes) a link which is active has at least one active neighbouring link, to the average delay of the link itself. 
In Fig.~\ref{fig:delays}B we show this quantity as a function of the average delay for both the Italian and German cases. 
We notice a slight increase in average delay as the co-activity increases, confirmed also by the Spearman's coefficients, $0.43$ for Italy and $0.56$ for Germany. 
Hence, the possibility of interaction between trains  in a certain part of the railway network seems to increase the delay localized in that area. 
Note that we have defined the \emph{first order co-activity} between neighbouring links, i.e. links that have at least one node in common. 
We can define a \emph{$k$-order co-activity} considering links connecting at least two nodes that are less than $(k-1)$ links apart following the shortest-path connecting them. 
We report the same measurements for $k=2$ and $k=3$ showing that in general, the same relation with the average delay is confirmed even though the curves are shifted towards lower values. 
This indicates that considering the interaction between non-neighbouring links is relevant but might include less important contributions to the delay. 
Hence, in the following we will limit our analysis to the first order neighbour links and assume that interactions between trains are possible only when they are in nearby links. 
Fig.~\ref{fig:delays}B supports the thesis of a propagation effect but an important feature has yet to be determined. 
In fact the direction of the propagation still has to be determined.

\subsection{Defining possible interactions}
\begin{figure}[t]
	\begin{center}
		\includegraphics[width=.6\textwidth]{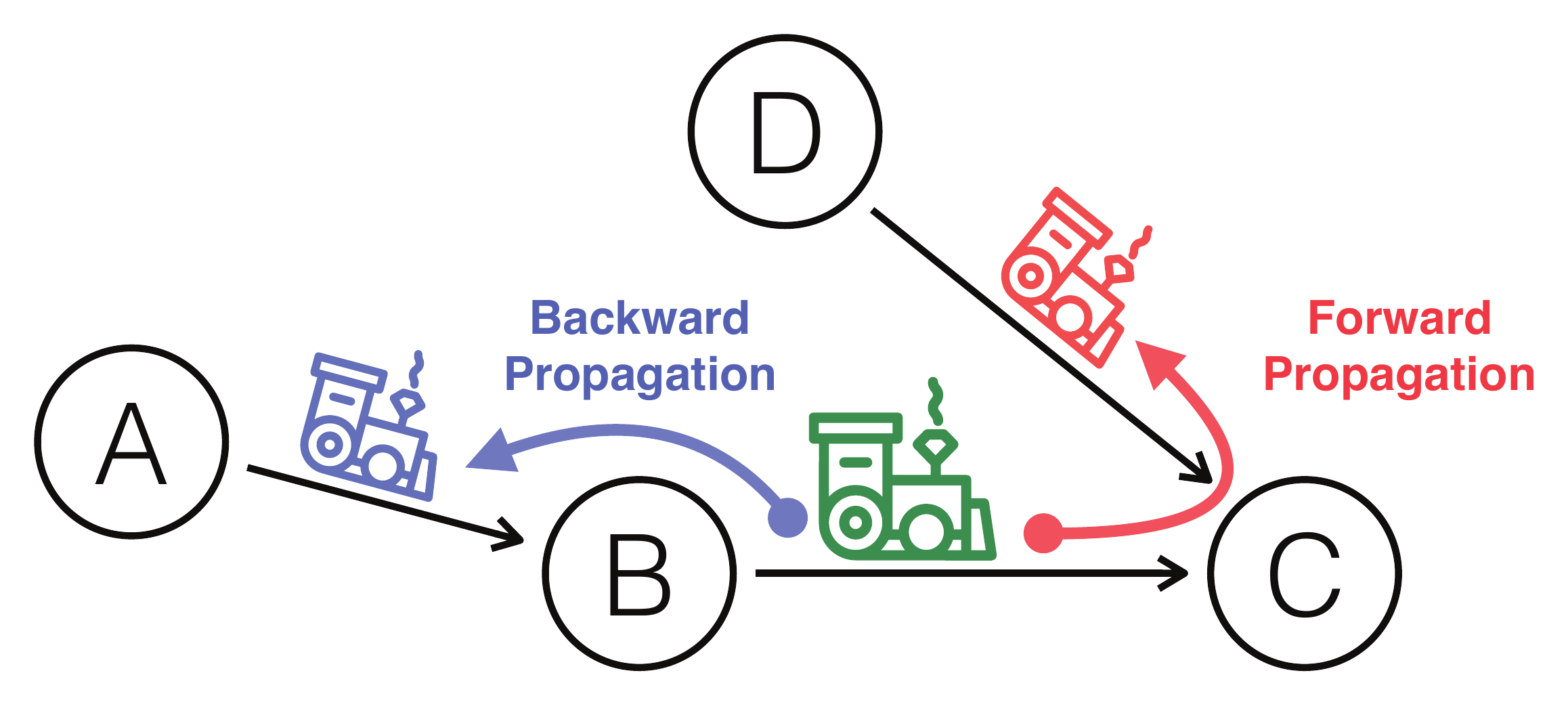}
		\caption{
			Arrows represent main delay propagation mechanisms.
			\emph{Backward propagation}: the (green) train travelling from station A to station B has got a delay and propagates it to following (blue) trains travelling towards station A;
			\emph{Forward propagation}: the train on AB propagates its delay to (red) trains that are travelling from the station C toward B.
		}
		\label{fig:propagation}
	\end{center}
\end{figure}

Let us consider a train $i$ travelling between two stations $A$ and $B$ (i.e., on the link $e=AB$, see Fig.~\ref{fig:propagation}) with some delay. 
We can argue that the propagation of this delay to other trains can occur only if they travel on the neighbour links of $e$. 
Due to the fact that the railway networks are directed, there are four different configurations of the links with respect to $e$:
\begin{itemize}
	\item[(i)]  links entering $A$, i.e. trains moving towards the last station crossed by $i$;
	\item[(ii)] links entering $B$, i.e. trains travelling towards the same station $i$ is currently travelling to;
	\item[(iii)] links exiting from $A$, i.e. trains departed from the last station crossed by $i$;
	\item[(iv)] links exiting from $B$, i.e. trains leaving the station $i$ is currently travelling to.
\end{itemize}
We can exclude the last two case: (iii) given that all the trains in such configuration will have no interaction with $i$; (iv) is less important because it describes scheduled connections. In the latter case, schedules foresee extra-time between the two trains exactly to avoid delay propagation.
We checked whether the propagation occurs in the case (i) of \emph{backward propagation}, in the case (ii) of \emph{forward propagation}, whose definitions are depicted in the graphic Fig.~\ref{fig:propagation}.
To discriminate which mechanism is at play, we measured the average delay time sequence $\Delta t (e)$ of each link $e$ of the network, defined as the average delay of all the trains that are currently travelling on $e$. Successively we measured the cross-correlation functions of the average delay time series of all the pairs of links, i.e., 
$\mathrm{CC}_{e,e'}(dt)=\sum_t \Delta t_{e}(t) \Delta t_{e'}(t+dt)/\sigma_{e}\sigma_{e'}$ 
being $e$ and $e'$ generic neighbours links of the network, $\sigma_{e}$ and $\sigma_{e'}$ the standard deviation of the whole time series $ \Delta t_{e}(t)$ and $ \Delta t_{e'}(t)$.
Then we averaged, aggregating the pairs of neighbours links according to their configuration (forward, backward, etc). In this way, for each of the four configuration, we obtained an average cross-correlation function.  In the backward propagation configuration can be defined :
\begin{equation}
\mathrm{C}_{\cal{B}}(dt)= \langle \mathrm{CC}_{e,e'} (dt)\rangle_{(e,e') \in \cal{B}}
\label{eq:crosscorr}
\end{equation}
with $\cal{B}$ as the ensemble of links pairs.
For both networks  the Backward mechanism is dominating while the Forward can be neglected (Fig.~\ref{fig:delays}D ).
The  high-speed layer of the railway network shows the similar backward mechanism, while there is no cross-correlations between the delays of high-speed vs. low-speed (see Supplementary Material Fig.~S1) acting as two independent layers. 
\subsection{Exogenous generation of delay}
\begin{figure}[t]
	\begin{center}
		\includegraphics[width=\textwidth]{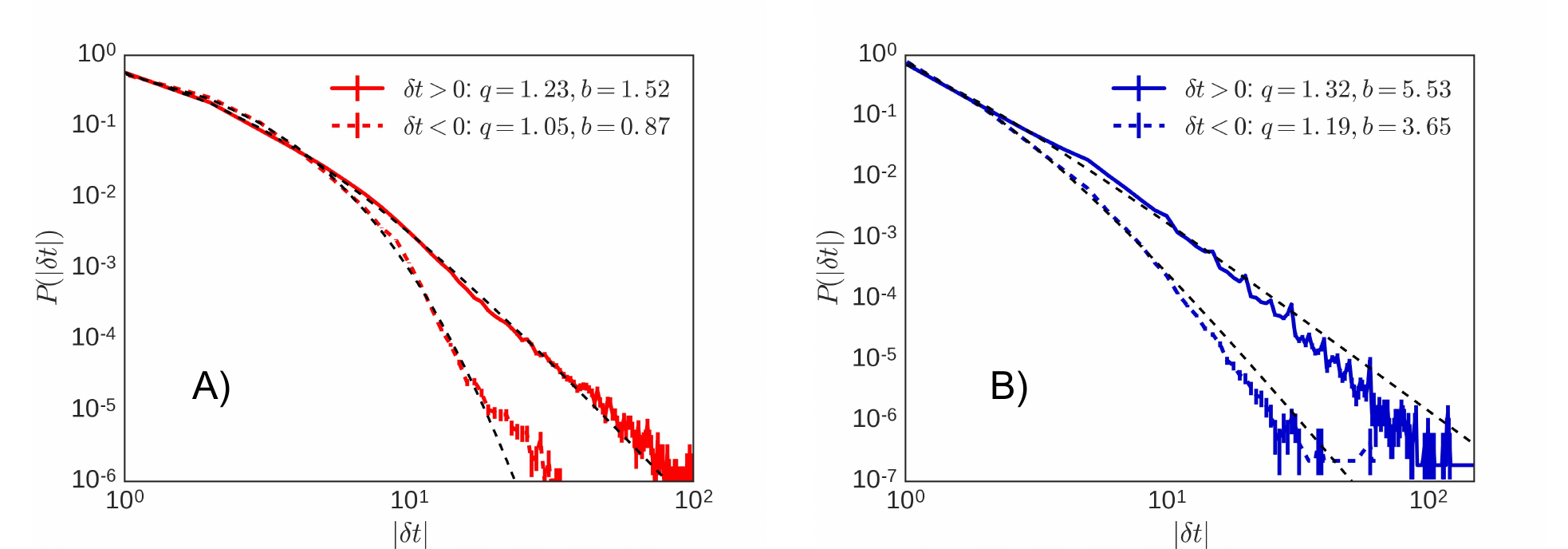}
		\caption{
			Exogenous delays: positive exogenous delays (continuous lines) and negative delays (dashed lines)  for the Italian (A) and German (B) railway networks. The black dashed lines are obtained by fitting the $q$-exponential distribution of Eq.~(\ref{eq:qexp}) to the data. The best fitting parameters are shown in the legend.}
		\label{fig:exodelays}
	\end{center}
\end{figure}

We define two kinds of delays: endogenous and exogenous.
By ``endogenous'' we mean that its origin is inside the railway system dynamics, i.e. it has been caused by another train. 
Conversely, by ``exogenous'' we mean that its cause is of a different nature: strikes, malfunctioning, bad weather or anything else which is not the result of the interaction with another train. 
We measure directly this types of delay in our datasets. 
Let us consider a train $i$ travelling from a station $A$ to a station $B$ on the link $e$ and further to a station $C$ on the link $e'$. 
It will travel first on the link $e$ and then on the link $e'$. The delay are indicate respectively as $\Delta t_i(e)$ and $\Delta t_i(e')$ . 
If there is a increase in the delay $\Delta t_i(e') > \Delta t_i(e)$ it might be ``endogenous'' or ``exogenous''. 
The exogenous delay is defined as $\delta t =\Delta t_i(e') - \Delta t_i(e)$, the variation of the train delay traveling on links whose neighbouring links were empty or hosted trains perfectly on time. 
It is worth noticing that $\delta t$ might also be negative, for example, if the train managed to make up for lateness. 
Results are reported in Fig.~\ref{fig:exodelays}, showing the distribution of positive exogenous delays as well as negative ones for the Italian and German cases.

\begin{figure}[t]
	\begin{center}
		\includegraphics[width=\textwidth]{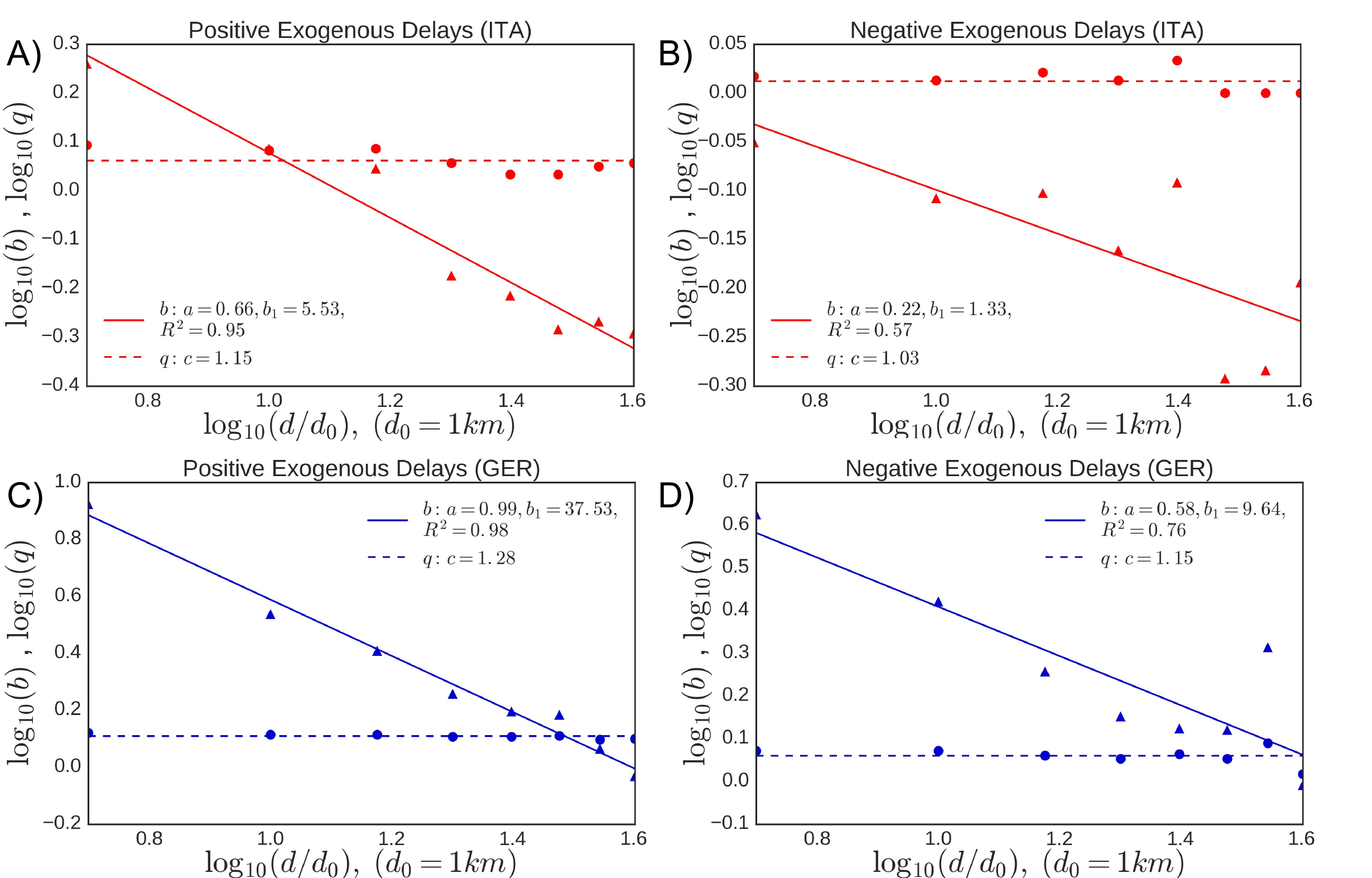}
		\caption{Best fits of the $q$-distribution parameters for the exogenous positive and negative delay distributions as function of the length of the links $d$. Panels A and B show the parameters for the positive and negative exogenous delays in the Italian railway network, while panels C and D show the same results for the German case. The function expressing $b$ was sought of the form $b=b_1 (d/d_0)^{-a}$, while the parameter $q$ of the constant form $q=c$.}
		\label{fig:exoparams}
	\end{center}
\end{figure}

In order to model these distributions, we adopted the same approach used in \cite{Briggs2007modelling} for departure delays. We fitted both the positive and negative parts of the distributions with $q$-exponential functions, where the parameter $q$ modulates from an exponential distribution $q\rightarrow 1$ to a fat-tailed distribution for $q\in (1,2]$\cite{picoli2009q}:
\begin{equation}
e_{q,b}(\delta t) \propto (1 + b(q-1) \delta t)^{1/(1-q)}~~\mbox{with}~~q\in[1,2], b>0.
\label{eq:qexp}
\end{equation}
It has been shown that such distribution can be obtained starting from a poissonian process $p(\delta t| \alpha)  = \alpha e^{-\alpha t}$, where $\alpha$ is a random variable extracted from of $n$ independent gaussian random variables $X_i$ with $\langle X_i \rangle=0$ and $\langle X_i^2 \rangle\neq0$, so that $\alpha=\sum_{i=1}^n X_i^2$ \cite{Briggs2007modelling}. 
In this way it can be proven that $n = 2/(q-1) - 2$, i.e.\ the parameter $q$, is related to the number of random variables composing $\alpha$. The parameter $b$ is proportional to the average value of $\alpha$, so that large values of $b$ at fixed $q$ result in a distribution biased toward shorter delays. The parameter $q$ quantifies how much equation (\ref{eq:qexp}) deviates from being exponential, which is the case $q=1$. This model has already been applied to the departure delays in the UK railway system, showing that the value of $q$ where so that $4\leq n \leq 11$ and thus estimating the number of independent occurrences contributing to the delay. 
For the  positive exogenous delays in the Italian and German case respectively, we found $q=1.23$ and $q=1.32$, corresponding to $n\simeq 7$ and $n\simeq 4$. 
The negative part of the distribution is exponential-like for the Italian railway network and broader for the German, this outlines the delay recovery strategies in the second case. 
To characterize the effect of the spatial distance on the delay distribution, we subdivided the links $e$ of the railway networks in classes according to the geodesic distance $d(e)$. 
Fig.~\ref{fig:exoparams} shows the behavior of the $q$ and $b$ parameters of the $q$-exponential fit as functions of $d(e)$. The parameter $q$ remains constant in every case, while on the other hand the parameter $b$ decreases as $b \sim d^{-a}$. Fig.~F-I of Appendix show the different best fit for equation (\ref{eq:qexp}) as $d(e)$ varies.
This result suggest that while the causes of the delay remain the same, the distribution of disturbances gets closer to a power-law as the length of the links increases, this outlines a relation between link length and delay. 
Finally, we can assume that the occurrence of positive or negative exogenous delays on links, $P(\delta t>0|d(e))$ and  $P(\delta t<0|d(e))$, are not roughly constant with $d(e)$ and hence are not depending on the length of the links (Fig.~J of Appendix).
\subsection{Generation of delay at departure}
Departure delays, i.e. the delay a train acquires right before leaving the first station on its route, cannot be considered in principle completely exogenous. 
In other words, due to the fact that different trains in our datasets can actually be the same physical train (e.g., the same convoy travelling back and forth along the same path on the railway network -- this is denoted as ``rotation'' --), the delay at departure might suffer from the influence of the traffic. 
However, railway administrators envisage suitable time buffers at the endpoints of the paths of each train so that it is reasonable to assume, at least as crude approximation, that departure delay is exogenous in character. 
It has already been shown that this kind of delay can be described by a $q$-exponential distribution in \cite{Briggs2007modelling}. 
However, the dependence on the parameters of the obtained distributions with respect to the topological properties of the network has not been investigated yet. 
Following the same spirit of the previous paragraph for the exogenous delay on links, we divide the nodes in the network (the train stations), with respect to their out-degree. 
The out-degree $k_\mathrm{out}$ represents roughly the number of different railway lines starting from a certain stations and hence can be considered as a proxy for the complexity of the station itself. Once the nodes of the networks have been divided according to $k_\mathrm{out}$, we fitted these distributions using a $q$-exponential following the procedure defined in \cite{Briggs2007modelling} (see Fig.~L, Fig.~M and Fig.~N of Appendix). 
\newline
Fig.~\ref{fig:departureparams} shows the behaviour of the parameters $q$ and $b$ of the $q$-exponential distribution as functions of $k_\mathrm{out}$ for the positive and negative departure delays in the two considered railway networks. Negative departure delays were never reported in the German dataset and hence we assume they are not present. 
Despite the fact that better proxies for station complexity than $k_\mathrm{out}$ might exist (weighting each link with the actual number of railway lines on it is a valuable alternative example), it is possible to see that we have again a constant parameter $q$ indicating that the sources of delays can be assumed to be the same independently from the station, while on the other hand the parameter $b$ decreases exponentially with $k_\mathrm{out}$. 
The small value of $R^2$ in the case of Italy might reflect the above mentioned possibility of having a non negligible endogenous contribution to the departure delays because of train rotations. 
\newline
In Germany the departure delay can be considered constant and independent on the size of the station. In Italy the  $P(\delta t>0\,| k_\mathrm{out})$ rows linearly with $k_\mathrm{out}$ meaning that stations with high degree are generating larger disruptions in the network.
\begin{figure}[t]
	\begin{center}
		\includegraphics[width=\textwidth, height=100pt]{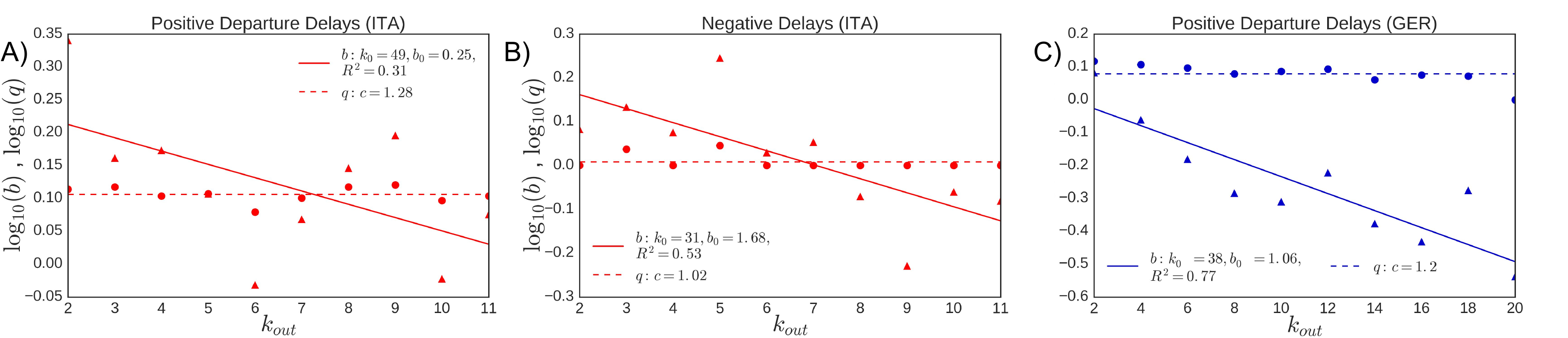}
		\caption{Best fits of the $q$-distribution parameters of the departure delay distribution as a function of the out-degree of the nodes $k_\mathrm{out}$. Panels A and B show the parameters for the positive and negative departure delays in the Italian railway network, while panel C the positive departure delay distribution for the German case. No negative departure delays were reported in the German dataset.
			The function expressing $b$ was sought of the form $b=b_0 e^{-k_\mathrm{out}/k_0}$, while the parameter $q$ of the constant form $q=c$.
		}
		\label{fig:departureparams}
	\end{center}
\end{figure}
\section{Modelling a realistic rail transport system}
\label{sec:modelling}
The analyses proposed so far suggest the existence of two sources of delays: \emph{exogenous delays} spontaneously occurring due to external adverse conditions at departure and during travel, depending on the topological properties of the Railway Network; \emph{endogenous delays} resulting from the interaction between trains. While in the first case we were able to characterize the statistical laws governing the emergence of delays, we cannot directly investigate the mechanisms of interaction between trains. Following the ideas exploited to model real world epidemics \cite{gomes2014assessing,zhang2016projected} which have already been proven effective on air traffic delay modeling \cite{fleurquin2013systemic}, we will define a propagation process for delays i.e. we will suppose that trains can spread their delays from one another according to some fixed probability and in certain conditions. Such probability will be derived by comparing the results from the simulations with the model and the empirical data.
\subsection{The Delay Propagation Model}
\begin{figure}[t]
	\begin{center}
		\includegraphics[width=.7\textwidth]{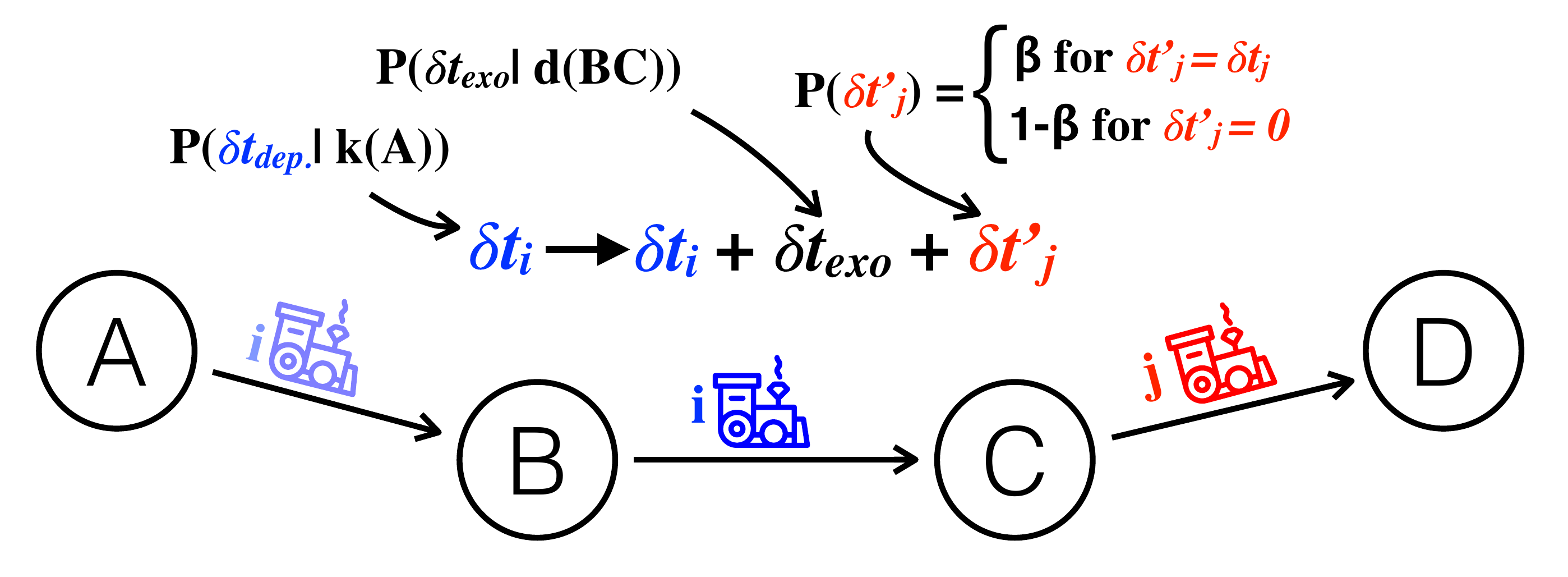}
		\caption{
			A schematic explanation of the core mechanism of the Delay Propagation Model. In this picture, a train $i$ departs from the station $A$ and acquires a departure delay from a distribution depending on the out-degree of $A$ in the railway network. 
			When the train $i$ passes through station $B$, its delay $\delta t_i$ gets two contributes: 
			i) a stochastic exogenous delay $\delta t_\mathrm{exo}$ depending on the length of the link $d$ the train $i$ is going to travel;
			ii) a propagated delay $\delta t_j$ depending on the presence of a delayed train $j$ travelling on a nearby link.
			If there is such a train $j$ with a positive delay $\delta t_j > 0$, this delay is propagated to train $i$ with probability $\beta$, which is the only free parameter of the model. 
		}
		\label{fig:model}
	\end{center}
\end{figure}

We define the model scheme by leveraging  the train-to-train propagation mechanism, with the aim of reproducing the real dynamics of delay spreading across the railroad network.
The model starts reproducing the normal schedule of trains on a certain day.
Train schedules are organized so that each train has its own ``time window'' to travel over a certain link along its path. 
Each train departure is at a fixed time and at a certain station and each intermediate station is going to be visited at a given time. However, our interest is in the deviance from the expected schedule. 
If some disruptions occurs a train might go out of the window of use for a certain link and overlap the window of another train for the same link. 
In this case, the second train will be forced to wait for its path to be cleared. 
Hence, the model adopts three different sources of delay as depicted in Fig.~\ref{fig:model}:
\begin{description}
	\item[\textbf{Departure delay}] 
	This delay is assigned at the beginning of the path (originating station) of each train and is considered exogenous and unrelated with the current traffic conditions at the departing stations or in the nearby links. 
	We assign to a train either a positive or negative delay according to the empirically found law of the corresponding probabilities $p_\mathrm{dep}^+=P(\delta t_\mathrm{dep} > 0\, | k_\mathrm{out})$ and  $p_\mathrm{dep}^-=P(\delta t_\mathrm{dep} < 0\, | k_\mathrm{out})$ respectively (and no delay with complementary probability $1 - p_\mathrm{dep}^+-p_\mathrm{dep}^-$) (see Fig.~O of Appendix).
	Once the sign of the delay has been decided, we assign a positive or negative delay value so that $|\delta t_\mathrm{dep}| \sim e_{q(k_\mathrm{out}), b(k_\mathrm{out})}(\delta t)$, i.e.\ distributed according to a $q$-exponential distribution with the parameters $q$ and $b$ depending on $k_\mathrm{out}$ and on the sign of the delay itself as in Fig.~\ref{fig:departureparams}. 
	\newline
	\item[\textbf{Exogenous Link Delay}] 
	 This delay is assigned whenever a train starts travelling on a link for the first time (Fig.~\ref{fig:model} reports an exemplification of the model). Considering a train $i$ passing from the link $AB$ to the link $BC$, we adopt the same modelling scheme as in the departure delay case assigning to the train a positive delay with probability $p_\mathrm{exo}^+=P(\delta t_\mathrm{exo} > 0\, | d(BC))$, a negative one with probability $p_\mathrm{exo}^-=P(\delta t_\mathrm{exo} < 0\, | d(BC))$ and no delay with $1 - p_\mathrm{exo}^+-p_\mathrm{exo}^-$, where $d(BC)$ is the geodesic length of the links $BC$ (see Fig.~J of Appendix for the corresponding law of probability). Having decided the sign of the delay, we fix the parameter of a $q$-exponential distribution $q$ and $b$ according to $d(BC)$ (Fig.~\ref{fig:exoparams}) and we extract the magnitude of the delay so that $|\delta t_\mathrm{exo}| \sim e_{q(d(BC)), b(d(BC))}(\delta t)$.
	\newline
	\item[\textbf{Delay Propagation}] 
	We model the interaction between trains with a mechanism of propagation of delay from other trains to train $i$. 
	The investigations performed on the datasets suggest that such interaction can occur only when nearby trains are in the \emph{Backward propagation} configuration of Fig.~\ref{fig:propagation}.
	\newline 
	Following the picture of the model in Fig.~\ref{fig:model}, when the train $i$ starts travelling on the link $BC$ leading to the $C$ station is susceptible of propagation from the train $j$, which is travelling from $C$ towards $D$. Inspired by the SIS models of epidemic spreading \cite{pastor2001epidemic}, in which agents can get infected and recover from the illness with a fixed probability, we introduce the delay propagation parameter $\beta \in [0,1]$ describing the probability that the delay of the train $j$ propagates to $i$. 
	We assume that the propagation occurs at the time $i$ starts travelling onto $BC$. 
	When train $i$ travels on $BC$ from time $t_1$ to time $t_2$, we check whether delayed trains are travelling on links in the Backward propagation configuration with respect to $BC$ in $[t_1,t_2]$. 
	Thus, we randomly pick one of those trains, say $j$, and with probability $\beta$ its delay $\delta t_j$ is added to the delay of $i$. 
\end{description}
The model simulates the theoretic schedule applying all these delay-generating mechanisms. 
More in detail, we start the dynamics of each train $i$ by adding a departure delay $\delta t_\mathrm{dep}$ according to the \textbf{Departure delay} mechanism. After the departure, each time the train starts traveling over a link in its route, we update its delay according to the following rule:
\begin{equation}
\delta t_i \rightarrow \delta t_i + \delta t_\mathrm{exo} + \delta t_j
\end{equation}
where  $\delta t_\mathrm{exo}$ is sample following \textbf{Exogenous Link Delay} mechanism, so that the delay will be positive with probability $p_\mathrm{exo}^+$ and negative with $p_\mathrm{exo}^-$ and the parameters of the sampling distributions are depending on the length of the link as in Fig.~\ref{fig:exoparams}. The term $\delta t_j$ the contribution of the Delay Propagation which will different from $0$ with probability $\beta$ and just if there is at least one delayed train in the right configuration.
This mechanism for intermediate stations reproduces the general noise associated with external causes, but does not account for long correlations such those present in case of large scale adverse conditions, e.g., bad weather, or national strikes. Finally, when the train reaches its final destination it is simply removed from the simulation. We used the described model to reproduce the delay spreading dynamics starting from a theoretic timetable and local exogenous delays distribution. Results of the simulation for both the Italian and German national railways systems are reported in the next section. 
\section{Results}
The model is based on the exogenous sources of delay that represent the spontaneous emergence of disruptions due to the aggregation of a finite number of external causes (trains malfunctions, accidents, bad weather, etc.).
These sources are modelled according to a universal probabilistic law, whose parameters are inferred from the data at our disposal.
The propagation parameter $\beta$ modulates the mutual interaction between trains. For simplicity, we chose this parameter to be uniform all over the network, which is a strong assumption since the propagation might depend on the considered part of the railway system.

\subsection{Reproducing the emergence of delays and congestion}
\begin{figure}[tbp]
	\begin{center}
		\includegraphics[width=\textwidth]{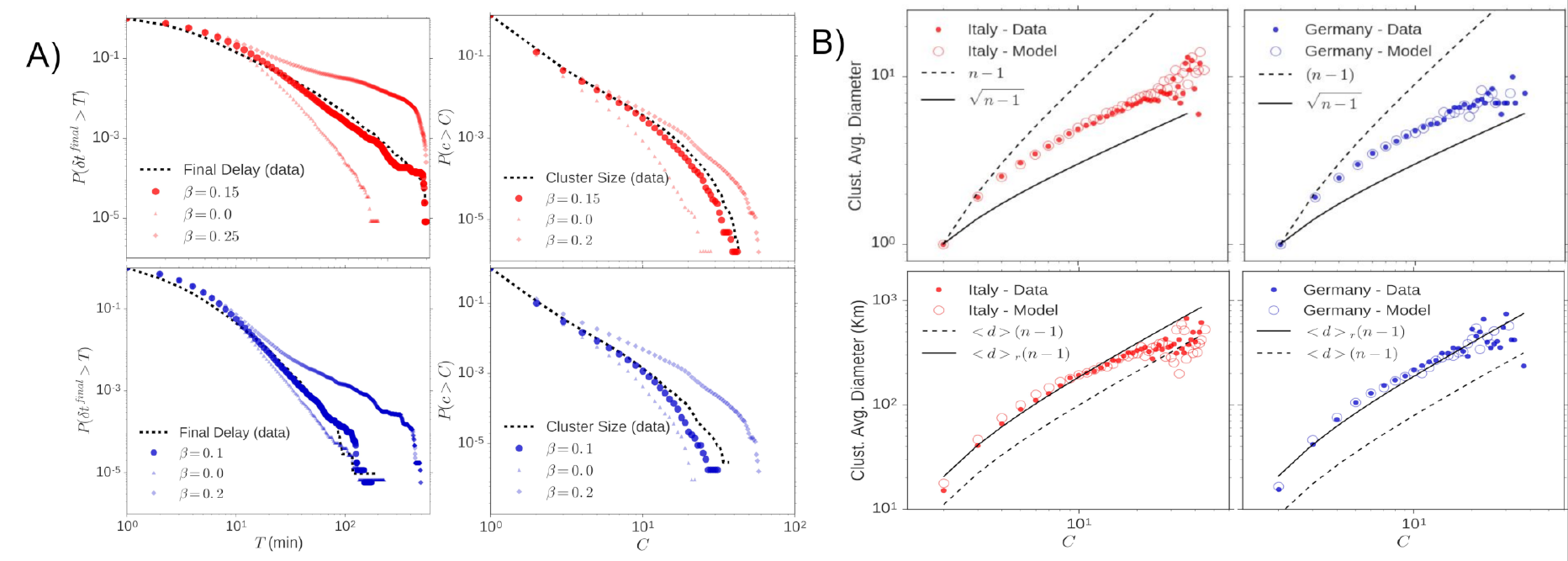}
		\caption{
			(A) Comparison between the empirical delay distribution (left) and the empirical cluster size distribution (right), with the results coming from the simulations for the Italian (Red) and German (Blue) Railway Networks. Different curves correspond to different parameters of $\beta$. The optimal $\beta$ value is $0.15$ for Italy and $0.1$ for Germany. The distribution shown in these panels are inverse cumulative distributions.			
			(B) Cluster Diameter as a function of the cluster size obtained with the empirical data and with the simulations. Dotted lines represent the case in which clusters are path-shaped.
			Left part represents results from data and simulation for the Italian system. Right part is for Germany. Upper panels show the diameter computed without considering the geographical distance between the nodes (with the black lines showing the behaviour in which clusters have a compact shape), lower panels take this distance into account.
			Results in A and B have been obtained by simulating the full-day schedules of March and April 2015, both for the Italian and German case. In this figure all the quantities that have not a specified unit of measurement are adimensional.
		}
		\label{fig:results1}
	\end{center}
\end{figure}

In the Appendix we show that it is possible to estimate the best value for $\beta$ by trying to reproduce the average delay of each station. These values have been estimated as $\beta=0.15$ for Italy and $\beta=0.1$ for Germany. In the top pabels of Fig.\ref{fig:results1}A, we show that with this choices for $\beta$ the system is capable of reproducing the empirical distributions of arrival delays. However, in order to disentangle the effect of the $\beta$ parameter and the two other sources of exogenous delay, we also report the same result for the case in which $\beta=0$ and $\beta$ is larger than the optimal value. In the latter, we clearly see that there is an increase in the number of extremely large delays. On the other hand, turning off the interaction by setting $\beta=0$ completely removes delays larger than $2$ hours. Hence according to our model, the presence of exogenous delays by themselves would not be able to predict the presence of large disruptions.
Since $\beta$ represents the probability of delay propagation between two trains, these estimations give a quantitative account of the difference between the two countries. 
These results suggest that the Italian trains propagate each other delay more often than German trains. 
The microscopic reasons of this difference could be connected to different properties of the railway networks, to the peculiar geographical structure of the territory, to different delay handling policies, etc., and it is out of the scope of the present paper.
As it happens in other transportation systems, we expect that disruptions occur clustered in certain areas of the network \cite{fleurquin2013systemic}. 
For this purpose we provide a definition of ``cluster of congested stations''  and how to discriminate whether a station is ``congested'', i.e.\ when its functioning is inefficient because of the hoarding of delays. 
For each station we can define a threshold between the ``functioning'' and the ``congested'' status based on the value of incoming train delays. 
We calculated such threshold for each station as the average value of the delays of all the trains arriving at the considered station in the dataset.
Considering a period of two months (March and April 2015), we examined each station every 5 minutes during the day, checking whether the average delay of all the trains moving towards that station during that lapse of time was above the average. 
We consider the station as ``congested'', while if the average delay is below that threshold we consider it as ``functioning''. 
\newline 
We identify, at each time step, the ``clusters of congested stations'' as the connected components of the railway network left  once we removed all the functioning stations from the network. 
(i.e.\ the stations whose congestion might be correlated with one another because of the propagation of delay).
\newline
In order to check whether our model is able to reproduce the emergence of these areas, we focused on two measures: 
(i) the size of the clusters in terms of number of stations and 
(ii) the relation between the size and the diameter of the clusters.
This latter measure is capable of giving insights about the topology of the congested clusters.
Bottom panels of Fig.~\ref{fig:results1}A show real and simulated cumulative distributions of cluster sizes for both countries. 
We find a strong accordance between model and reality when $\beta$ is set to the optimal value, pointing out how the model is also able to reproduce this aspect of delays dynamics. 
Similarly to the arrival delay distribution, here $\beta$ affects the tail of the distribution so that $\beta=0$ (i.e. no propagation of endogenous delays) implies the absence of large congested areas.
We study the diameter of the cluster defined as the distance between the farthest couple of nodes in the cluster. 
We compute the diameter both in the case where just the topological distance between nodes is considered and the case where each link is weighted with the geodesic distance between the stations it connects. 
To panels of Fig.~\ref{fig:results1}B shows the dependence of the cluster diameter computed with both distances on the cluster size. 
In the cases where the diameter is computed by using the geodesic distance, the dotted line represents the diameter of a cluster of size $n$, assuming that all the links of the clusters are as long as the average link length of the network. 
Such pattern corresponds to the case where all the clusters are randomly sampled from the whole network without any constraint. 
From the lower panels of Fig.~\ref{fig:results1}B we can see that while clusters do have a path-like structure, they cannot be considered randomly distributed over the networks.  
Instead, they seem to be deployed in areas where the links are larger than the average links length of the corresponding network, resulting in the same dependence of the dotted lines but shifted upwards. 
The fact that the geographical diameter of large clusters grows up to hundreds of kilometers indicates that disturbances can propagate between far away parts of the network.
\newline
The topological measure in the top panels of Fig.~\ref{fig:results1}B exhibits strong deviations from the path-like behaviour especially when large clusters are involved.
The mismatch between the two ways of characterizing clusters could be due to the fact that the geographical distance may hide, by stretching them, non path-like structures that are actually present in the delay propagation patterns. 
The appearance of these clusters, and their features, are typical outcomes of a non-trivial complex dynamics arising from train interactions. For this reason it is a remarkable result that, as can be neatly be seen in Fig.~\ref{fig:results1}, the model proposed, despite its simplicity, captures this behaviour correctly. The relations between cluster size and cluster diameter is show very good agreements with the empirical measures for both nations and, perhaps more importantly, for both clusters diameter definition: the topological and the spatial one. This clear adherence with reality of the results from a model with only one parameter trained on a so small training set (only one week) is a strong proof of the fact that the model hypothesis are very likely to be correct.
Our model is capable of reproducing some global patterns of the delay dynamics in railway systems. However, some discrepancies at a more microscopic level can be observed. 
In particular our model fails to correctly describe the behaviour of stations with low traffic. For these stations, the hypothesis of a constant in time and homogeneous coupling parameter $\beta$ may not be well justified. We refer to Section~5 of Appendix for a thorough discussion of this point.
\subsection{Scenario Simulation: Prediction of the Effects of a Strong Localized Disturbance}
\begin{figure}[htbp]
	\begin{center}
		\includegraphics[width=1.1\textwidth]{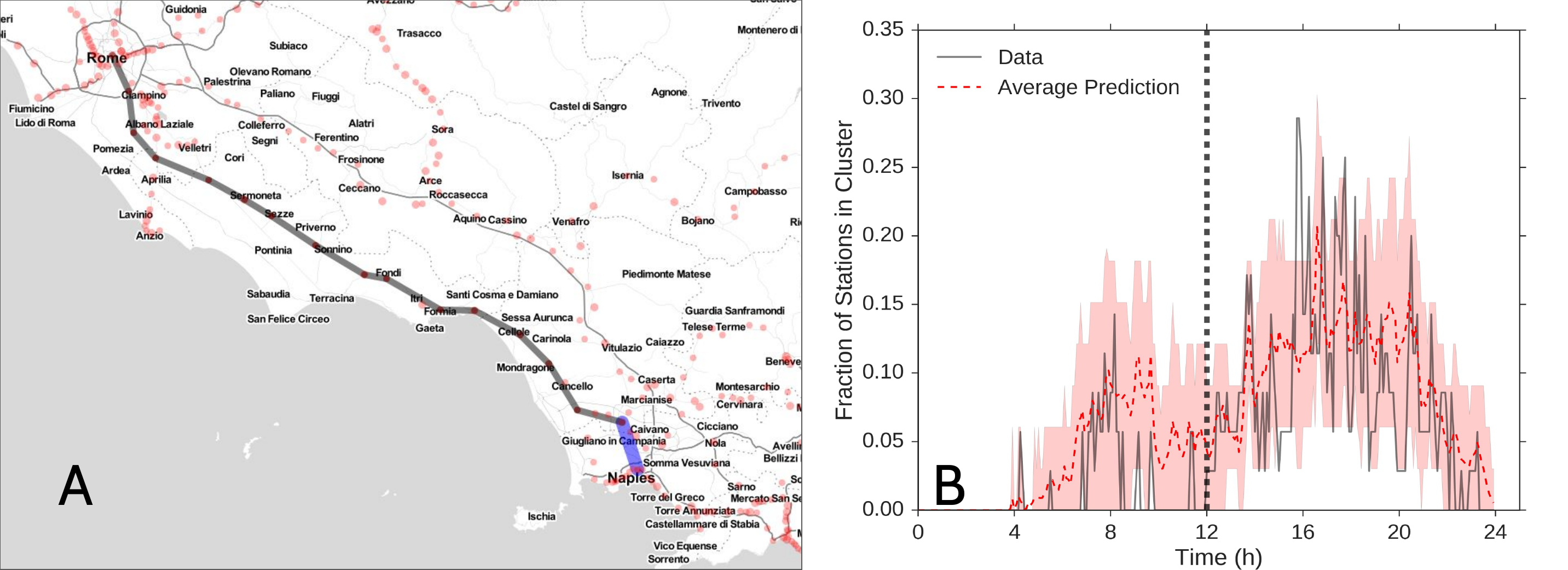}
		\caption{
			(A) Representation of the ``NAPOLI CENTRALE - ROMA TERMINI'' route. The highlighted blue link is the ``NAPOLI CENTRALE-AVERSA'' one, i.e.\ the spark of the perturbation. 
			(B) Fraction of congested stations in the ``NAPOLI CENTRALE - ROMA TERMINI'' route at different times of the day. 
			The solid black line represents the empirical measure performed on the dataset, the red area represents the results of the simulations between the 5th and 95th percentile of all the realizations. 
			The dashed red line is the average fraction of congested stations obtained by averaging over the realizations. 
			The vertical line represents the beginning of the perturbation on the ``NAPOLI CENTRALE-AVERSA'' link.
		}
		\label{fig:result_casestudy}
	\end{center}
\end{figure}

According to our model, the emergence of large clusters of congested stations is due to the propagation of delay between trains. The source of this delay is however exogenous, in the sense that comes from some adverse conditions which are external with respect to the interaction of the trains. 
In order to investigate how the emergence of a real large cluster is linked to exogenous effects, we study the case of the cluster emerged in the eastern part of the Italian Railway Network on the 28th of February 2015. 
As can be seen in the on-line visualization\footnote{The visualization can be found at \url{www.riccardodiclemente.com/trainsimulation.html}}, a large congestion starts to emerge in the early afternoon around 12am and propagates to a large part of the network until night. 
We empirically identified the interested part of the network as the shortest-path connecting the station of ``NAPOLI CENTRALE'' to the station of ``ROMA TERMINI'', indicated in Fig.~\ref{fig:result_casestudy}A. 
The shortest-path has been computed weighting each link with the geographical distance between the stations it connects. Fig.~\ref{fig:result_casestudy}B shows the fraction of stations in this path that are congested in different times of the day, clearly indicating that such fraction starts growing until a peak is reached in the afternoon. 
We have been able to identify the beginning of this adverse occurrence as a disruption on the ``NAPOLI CENTRALE-AVERSA'' link (highlighted in Fig.~\ref{fig:result_casestudy}A) in the first part of the afternoon from 11\,am to 14\,pm, resulting in a large delay acquired by the trains travelling on that links. 
We argued that this disruption was the spark that lightened the emergence of the congested cluster.
\newline
In order to check this hypothesis, we ran a simulation in which a large delay of $100$ minutes is assigned with probability $1$ to each of the trains crossing the ``NAPOLI CENTRALE-AVERSA''  link in that period of time. 
Due to the non-deterministic nature of the model, we performed the simulation of $200$ different realizations in order to have a set of scenarios to be compared with the real data. 
This comparison is shown in Fig.~\ref{fig:result_casestudy}B. 
We can see that with this simple modification and not considering other possible correlations between the occurrences of external adverse conditions in nearby links, we are able to reproduce a pattern which is qualitatively similar to the one observed in the data, clearly pointing out that the ``NAPOLI CENTRALE-AVERSA'' link played a major role in the congestion of the network. Moreover, our scenarios indicates the probability of having a minor congestion in the line also in the early morning, probably due to usual minor disruption occurring on other links.
The proposed scenario simulation could be easily extended to less localized adverse occurrences, which might comprehend spatially correlated disruptions due to natural events or strikes. Moreover, the same approach could be used to study more dramatic effect of node, link removal or the positive effects due to the introduction of more resilient and optimized schedules. In this cases, it is sufficient to use the same structure of the model and modify the input schedule and/or the Railway Network itself.
\section{Conclusions}
\label{sec:conclu}
Railways and the railway transport systems have been historically of utmost importance for the development of modern countries. Nowadays their importance is on the rise again due to their relevance in the reduction of CO$_2$ emissions and their competitiveness in short and middle range movements, so that huge investment have been made in the European Union to improve their efficiency. 
However, the emergence of large disruptions and the intrinsic inefficiencies seem to be endemic in this kind of transportation. 
The understanding of the universal features governing the dynamics of the system from a theoretical point of view could give an important contribution to the solution of such problems. The development of a universal model explaining the emergence of delays in Railway Network would allow for a quantification of the causes behind the occurrence of large disruptions and for a more direct link between the effects that localized interventions might have on the global system.
In this work, we developed a novel model of delay emergence and propagation between trains moving on Railway Networks, partly based on empirical laws inferred from the data governing the accidental emergence of delays from the influence of adverse occurrences.
Remarkably, the statistical models used for the description of this ``exogenous delays'' showed how these adverse occurrences are the result of the same finite number of causes  (e.g., bad weather conditions and malfunctions) independently from the topology, as opposed to the scale of these disruptions, which are largely influenced by the part of the network the trains are travelling on.
We find, in fact, that both the complexity of a station as measured by the number of different routes originating from it and the length of a route, are connected to the magnitude of the microscopic delays composing the overall delay by simple statistical laws with two parameters, whose value is the same all over the network and can be inferred and fixed by data. 
This kind of universality is somehow surprising  and opens the possibility of easily simulating the behaviour of \emph{any} railway system once the complete schedules are known.
These emergent delays can then spread from one train to another by means of a delay propagation probability $\beta$, which is in fact the only free parameter of our model. 
Despite the high level of abstraction used in our approach, our model is capable of reproducing the empirical patterns found in  data when the delay-spreading probability $\beta$ is fine tuned to an optimal value. 
The model reproduces correctly the final delay distribution and the distribution of the sizes of congested areas. Moreover, the model grasps the topological properties of such congested areas, which appears to be spreading in some cases for many kilometres through the network. According to our model, the emergence of large disruptions is the result of the interplay between the occurrence of localized exogenous delays and the propagation of such delays between trains. In other words, turning off the delay propagation mechanism prevents the system from generating extremely large delays and congested areas, pointing out that interactions are the driving force behind the emergence of major spontaneous adverse occurrences. 
The modelling scheme seems quite promising so far, but it still suffers from some major assumptions that might limit its predictive power. In fact, the interactions between trains could occur in longer ranges and not just between neighboring links and the propagation probability could not be uniform all over the Railway Network but instead depending on specific operational conditions. Moreover, the modelization of exogenous delays could be refined by taking into account more static topological feature of the Railway Networks or by introducing dynamical variations due to different traffic conditions. Another interesting possibility would be to increase the number of Railway Systems under study (e.g., the French SNFC sharing similar characteristics with respect to the Italian and German systems whose data is publicly available \cite{sncf}), in order to understand the reasons behind the quantitative differences observed, e.g. larger delays in the Italian case.
In the final part of the paper, we show how the model can be applied to study the capability of functioning of the system after a localized large disruption occurring in a single node of the network. This approach can be easily extended to more complex case studies of distributed disruptions, the occurrence of strikes, the removal of trains or parts of the network, also including the possibility to introduce delay management strategies to increase the resilience of the system. The model is also useful in order to have a fast test of changes in the overall schedule of trains so to have a more precise assessment of their global effects.  
Finally, despite its simplicity the model is open an increase of complexity that might lead to a better adherence to the empirical findings such as long-range interactions between trains and a more detailed model of exogenous delays including correlations between nearby link and the dependence on traffic conditions.
\section*{Funding}
This work has been supported by the KREYON Project, funded by the John Templeton Foundation under contract n. 51663; VDPS acknowledges financial support from the Austrian Research Promotion Agency FFG under grant $\#$857136. RDC as Newton International Fellow of the Royal Society acknowledges support from The Royal Society, The British Academy and the Academy of Medical Sciences (Newton International Fellowship, NF170505).
\section*{Acknowledgements}
The authors acknowledge Fabio Lamanna for the initial discussion about the datasets to be used for the work.

\FloatBarrier
\appendix
\setcounter{figure}{0}    
\setcounter{table}{0}    
\renewcommand{\thetable}{\Alph{table}}
\renewcommand{\thefigure}{\Alph{figure}}
\section*{Appendix}
\section{Datasets Information}
Novel information technologies enabled real-time monitoring and sharing of any kind of traffic data. 
Impressive instances are the visualised datasets about marine traffic that can be easily  found on the Internet\footnote{E.g.: \url{https://www.shipmap.org}, \url{https://www.marinetraffic.com}, \url{https://www.vesselfinder.com}}.
Also, several websites display live air traffic, by gathering and visualising official data from various sources\footnote{E.g.: \url{https://www.flightradar24.com}, \url{https://flightaware.com}, \url{https://planefinder.net}}. 
These sources of information were found to be crucial in order to improve the understanding of the related transportation systems \cite{guimera2004modeling,guimera2003structure,kaluza2010complex}.
However, it is still a hard task to aggregate and analyse global or continental datasets about railway systems. 
In fact, due to  historical reasons and to the typical usage scale, each nation has a network with few international connections and the available datasets are not homogeneous in coverage and format. 
Thus, tailoring the analysis on national systems seems a natural choice, whereas the most interesting characteristic behaviours appear in all systems, suggesting some kind of universality in the dynamics. 
We focused our analysis on the European continent both for the historical importance of railways and for the recent institutional efforts to raise their adoption. 
The actual railway system is composed by three distinct layers: high-speed passenger trains, normal-speed trains (mostly of regional type) and freight/military trains.
While the freight trains use different stations and traffic handling rules (e.g., they operate mainly during night time) and can be discarded from our analysis, the other two layers can possibly interact each other. 
In the high-speed layer, correlations are identical to those in the regular layer, while no appreciable correlations can be spotted across the two layers so that considering them as independent is a fairly good approximation.
Since no additional information can be gained by studying the whole system, we focus on the regular-speed layer only, both for Italy and Germany
\subsection{The Italian Dataset}

The dataset regarding the Italian Railways has been collected by means of the ``ViaggiaTreno" website\footnote{\url{http://www.viaggiatreno.it}}. 
The purpose of this website is to provide real-time information to travellers regarding the position of a certain train on the network, its delay and possible adverse occurrences like cancellations or strikes.
Despite the fact that the information is in real-time, i.e., the instantaneous delay of a train can be checked at any time during the day, whenever the train arrives at its final destination its record is not deleted from the site. 
Instead, it is possible to check its route and its delay at each intermediate stop from the departing station to the arrival one until the end of the day, at 23:59.
Hence, we downloaded all the information displayed on the website each day at 23:30 in order to be sure that each train would have arrived at destination.
Starting from the 1st of January 2015 and for the whole 2015, we collected 12 months of historical data about the dynamics of regular and high-speed trains in Italy. 
For each train we get an identifier, the ordered list of stations the train has to cross, the scheduled arrival time at each station and its delay. 
The resulting dataset comprehends the traffic running on $2253$ stations, with a daily average schedule pertaining $8112$ trains on $7062$ links. 
Note that ``ViaggiaTreno'' does not collect information about the geographical position of the stations. 
Such information has been integrated by means of Wikipedia and Google Maps, allowing us to represent the geo-localized network of Italian railways.
Each dot corresponds to a station and each link corresponds to a route between two stations. In other words, a line between Rome and Naples means that there is a direct train route linking them without intermediate stops. Lacking real point-wise tracks data, the route has been simply represented with a straight line.
\subsection{The German Dataset}
The data about German Railways have been collected through the OpenDataCity\footnote{\url{https://www.opendatacity.de}} website. This site gathers different datasets collected by a variety of on-going or terminated projects dealing with open data. 
In particular, the data we analysed come from the ``Zugmonitor'' project, which aimed at providing a web-app and an API to German travellers in order to have real time information on the position of the trains on the German Railway Network and their delay. 
The project is no-longer running and the API is not accessible anymore. 
However, some dumps of historical data collected during the project are still available. In particular, we downloaded all the data regarding year 2015, covering  the same period of the Italian dataset. 
This dumps collected not only the delay at each station like in the Italian case, but also the delay at intermediate points between two stations. 
All the points are also geo-localized so that it is possible to reconstruct a quite accurate trajectory of the trains. 
In order to be consistent with the Italian dataset we used the geo-localization only to identify the position of the stations in the map. Scheduled arrival times at each station were also stored in the dump, so that in the end we managed to reconstruct a dataset with a structure identical to the Italian one. 
The resulting dataset includes data for 5979 stations with a daily average schedule containing 11,975 trains on 16,277. 
\section{High-speed layer}
\begin{figure}
	\centering
	\includegraphics[width=150pt]{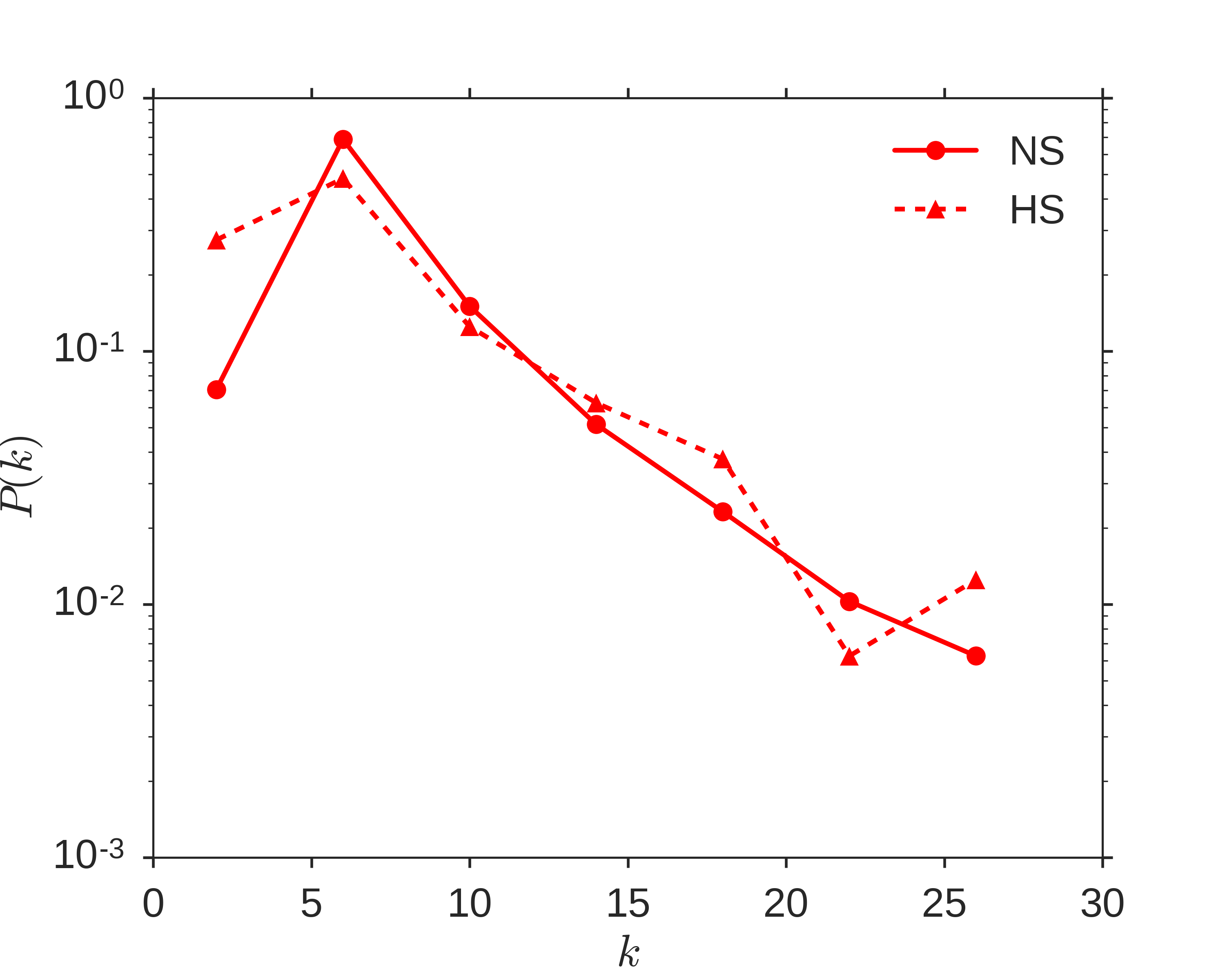}
	\includegraphics[width=150pt]{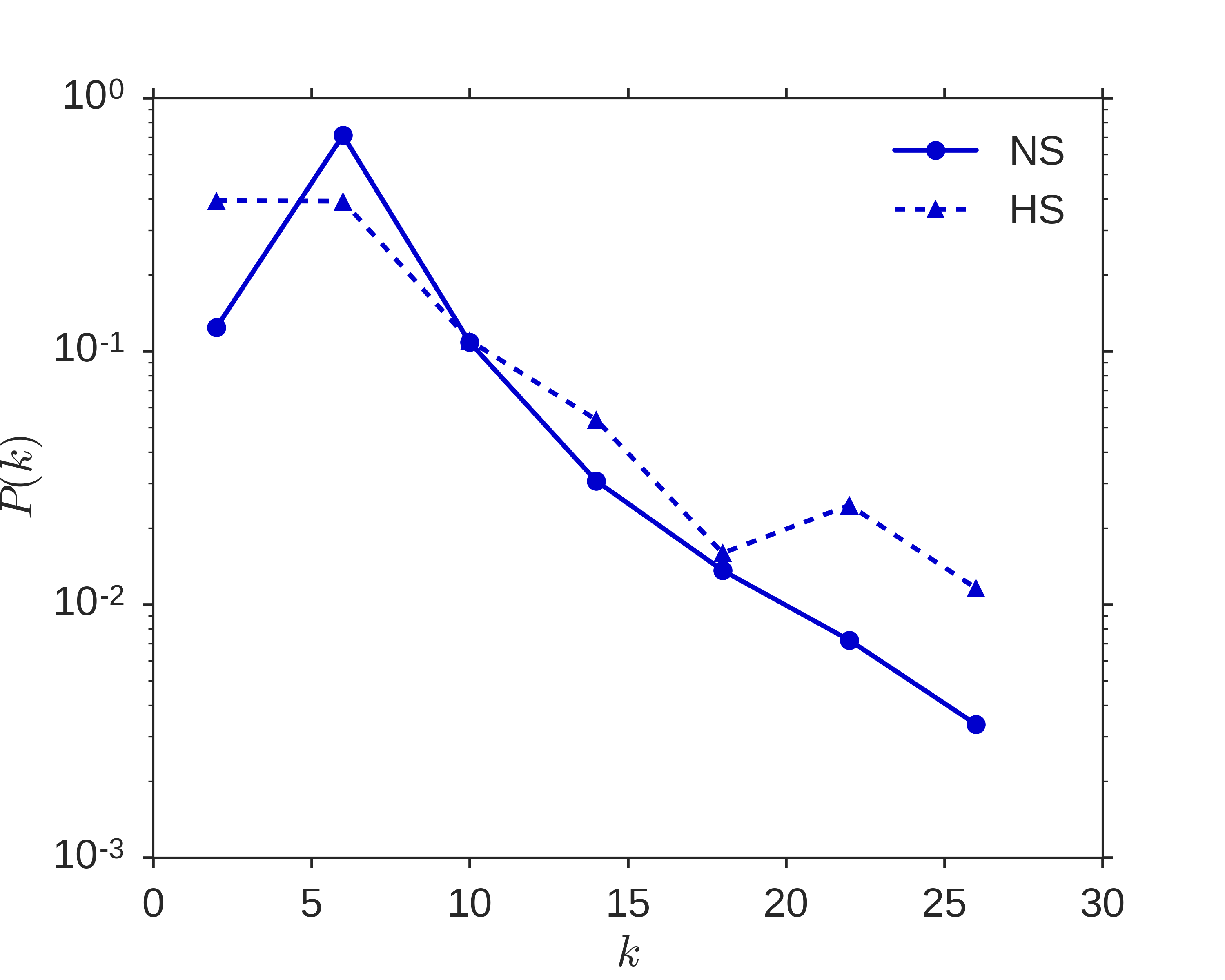}
	\caption{Degree distributions for the Italian (left) and German (right) Railway Networks. Continuous lines correspond to the normal-speed layers, while dotted lines correspond to the high-speed layers. }
	\label{fig_deg_dist}
\end{figure}

\begin{figure}
	\centering
	\includegraphics[width=150pt]{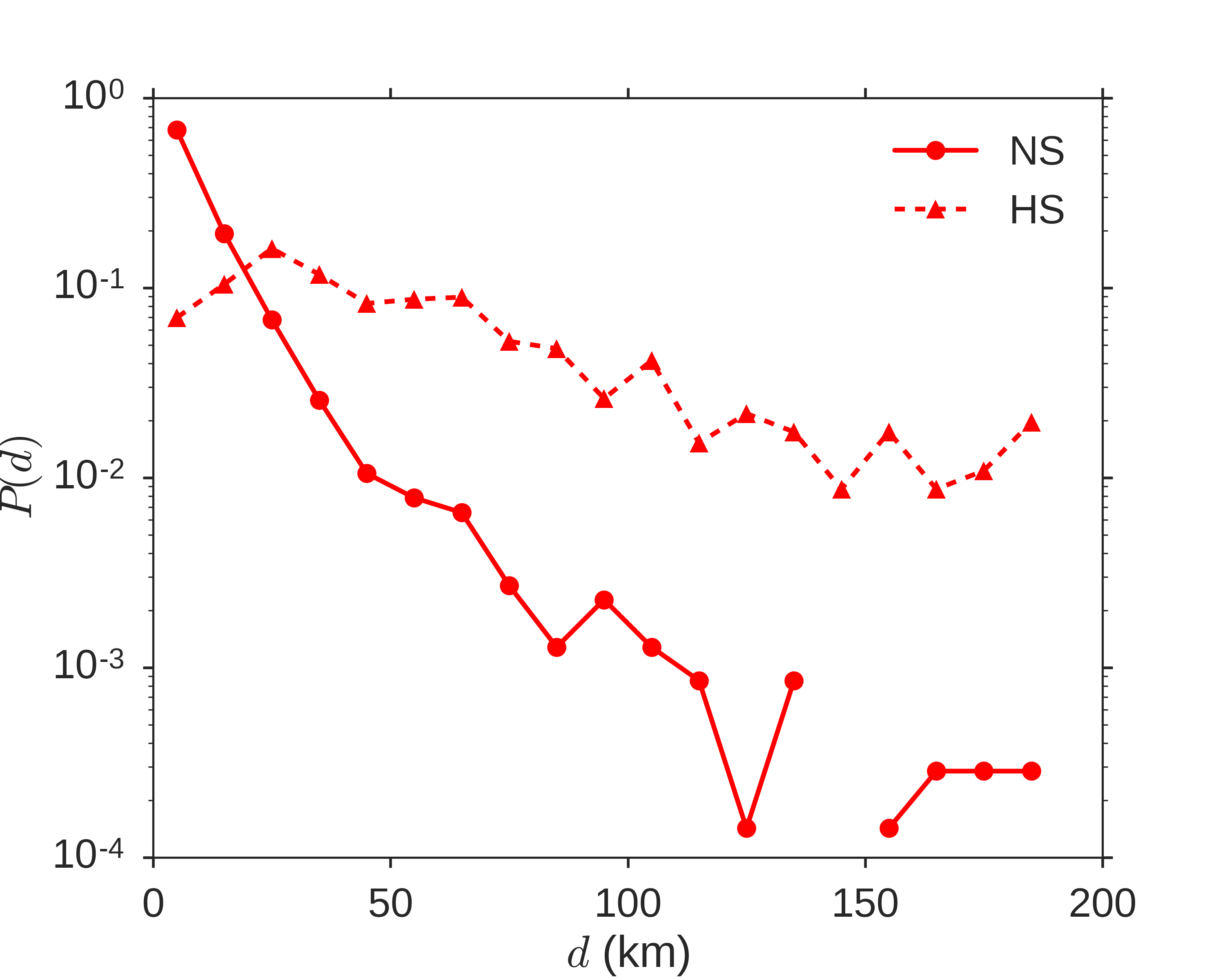}
	\includegraphics[width=150pt]{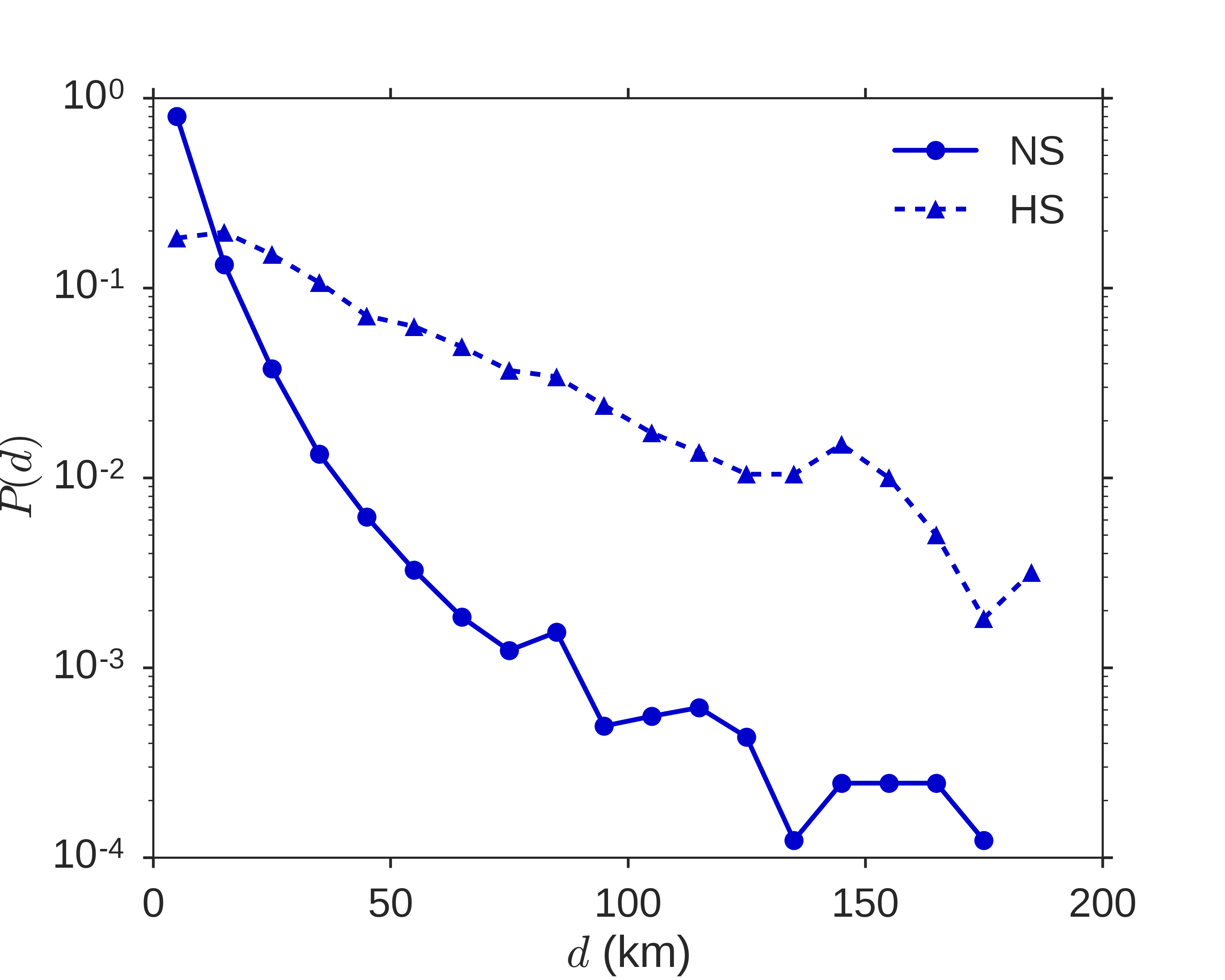}
	\caption{Links Length distributions for the Italian (left) and German (right) Railway Networks. Continuous lines correspond to the normal-speed layers, while dotted lines correspond to the high-speed layers. }
	\label{fig_distances_dist}
\end{figure}

\begin{figure}
	\centering
	\includegraphics[width=150pt]{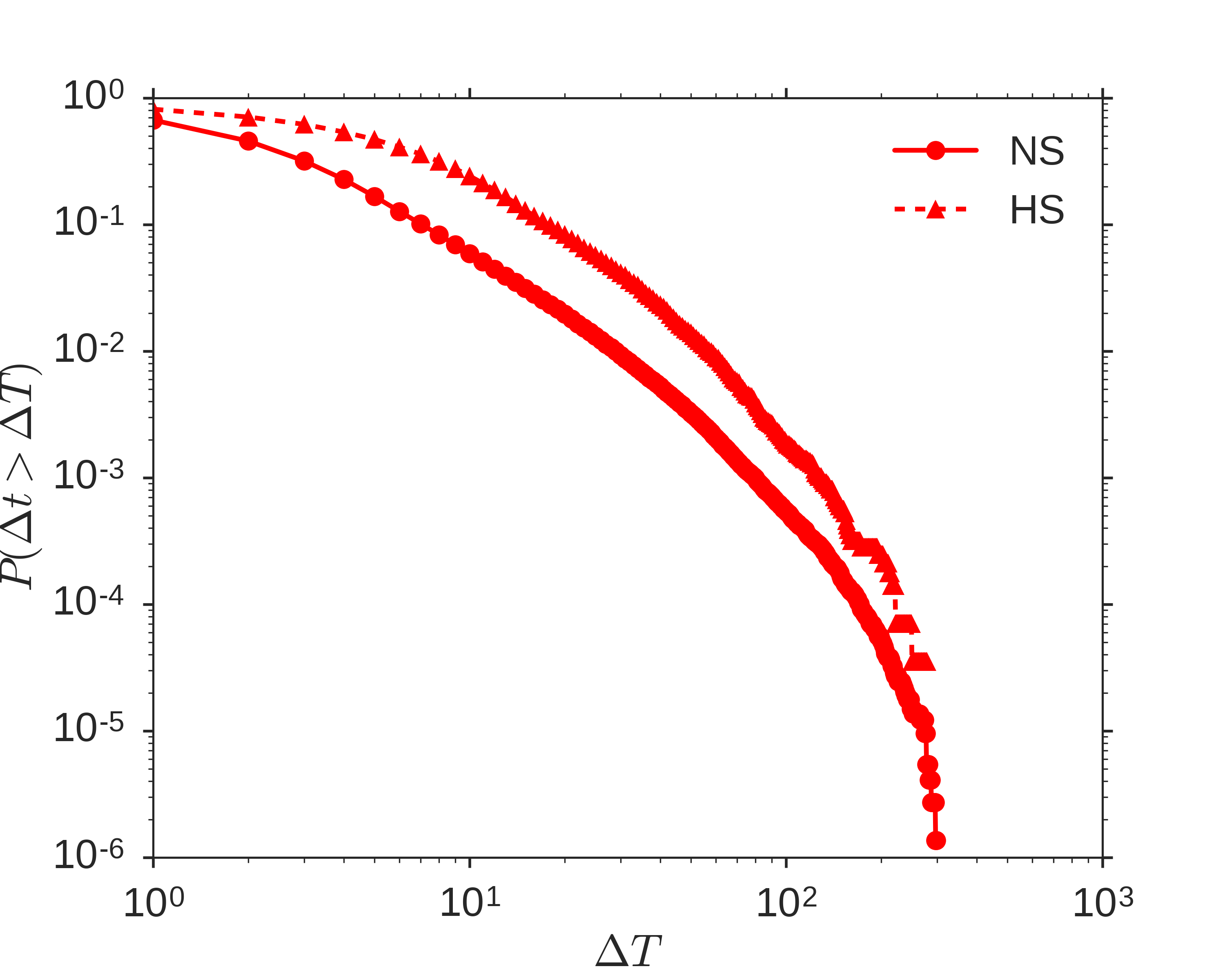}
	\includegraphics[width=150pt]{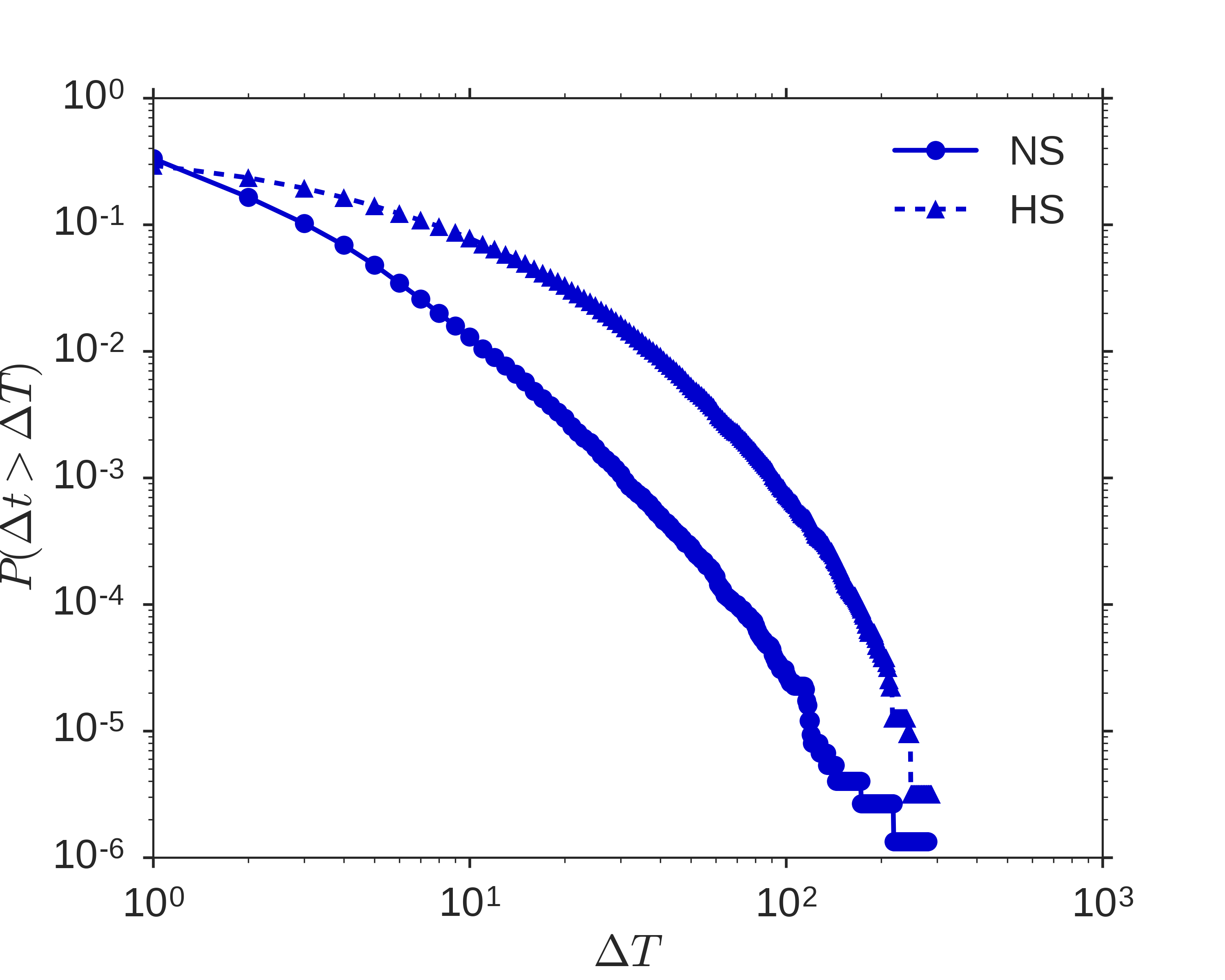}
	\caption{Final delay distributions for the Italian (left) and German (right) Railway Networks. Continuous lines correspond to the normal-speed layers, while dotted lines correspond to the high-speed layers. }
	\label{fig_final_delays}
\end{figure}

\begin{figure}
	\centering
	\includegraphics[width=150pt]{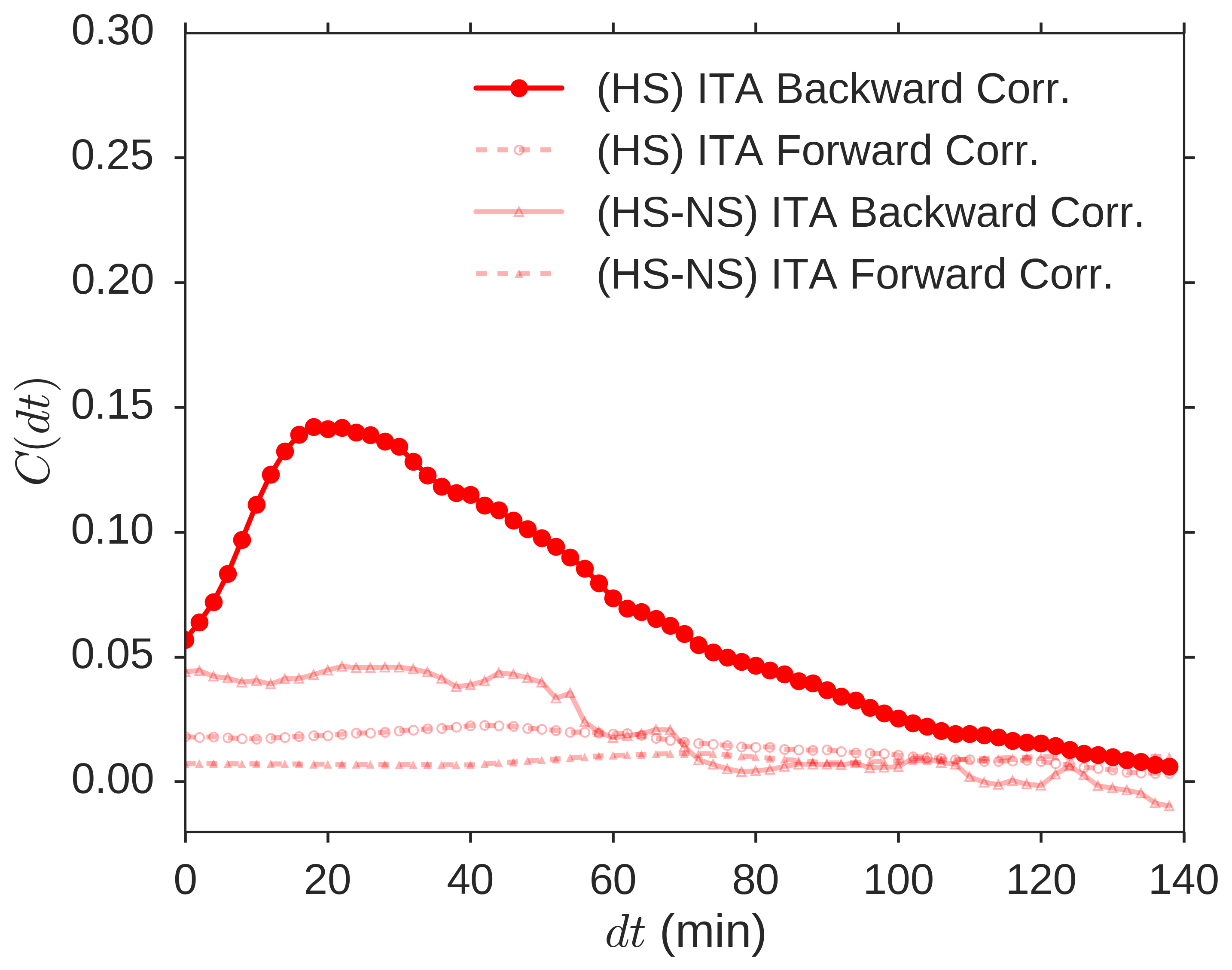}
	\includegraphics[width=150pt]{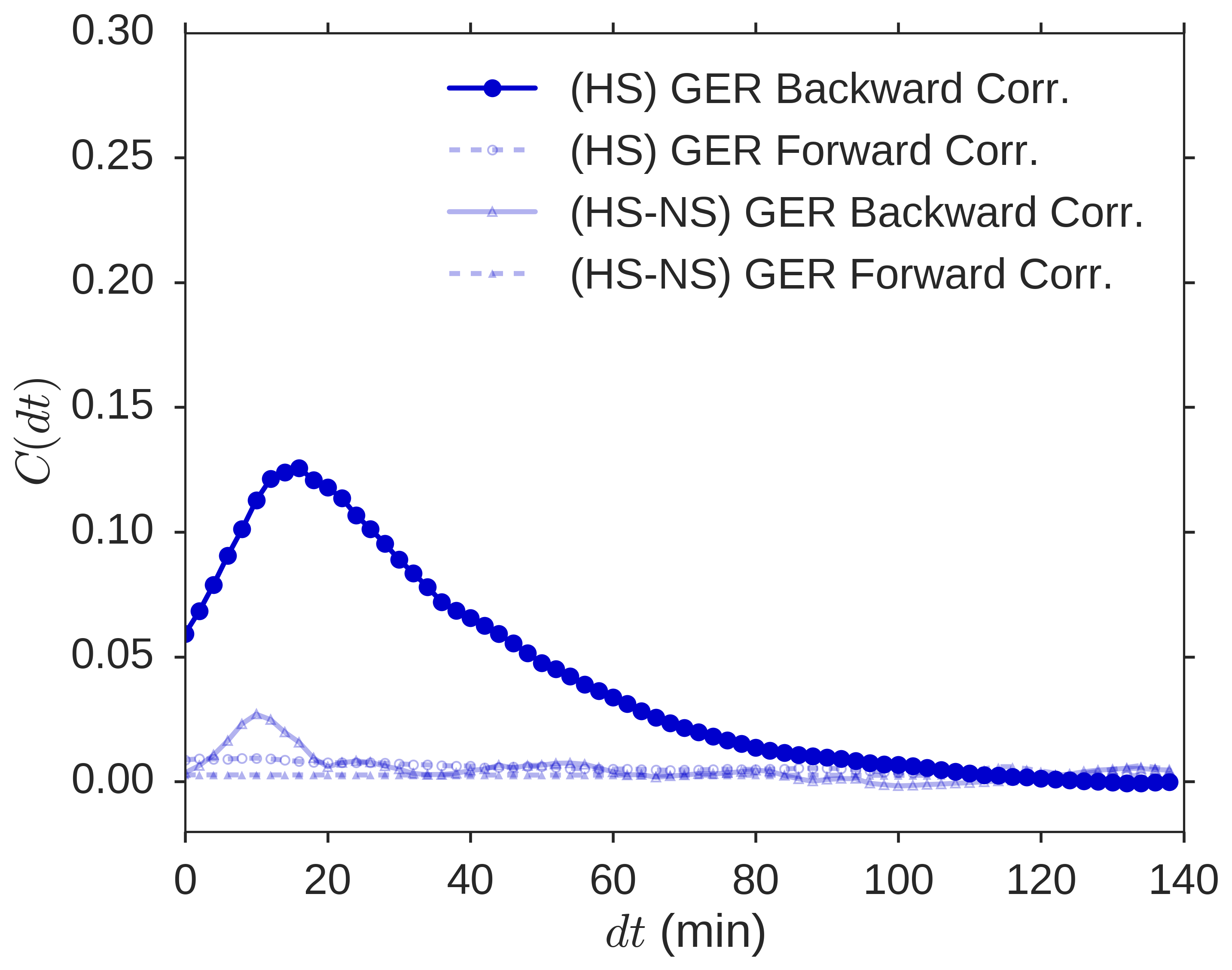}
	\caption{Cross-Correlations between the average delay time series of pairs of nearby edges in the high-speed layers of the railway networks and between the time series of pairs of edges coming from different layers. Decaying correlations are observed only in the ``Backward'' case for pairs of edges both in the high-speed layer. No signal of inter-layer correlations can be observed.}
	\label{fig_crosscorr}
\end{figure}

The structure Italian and the German Railway Networks is the overlap of two distinct layers, the normal-speed and the high-speed one. These two layers are different from the structural point of view. The high-speed layer in fact has to allow for fast travelling trains and have a different kind of rails connecting stations and, in general, it is reasonable to assume that high-speed trains and normal-speed ones do not interact when travelling from a station to another. However, the nodes of the network (i.e. stations) are shared between the layers, making the network a ``multiplex'' \cite{battiston2014structural,menichetti2014weighted} and allowing for interactions of the two different kind of trains. \newline Our datasets contain information about high-speed and normal-speed trains, for both the Italian and German case and in principle it could be possible to study the dynamics of the high-speed layer and its interaction with the normal one. In the main text we decided though to focus on the normal layer cutting-out the high-speed part. This choice was made for sake of simplicity, since the rules of interaction between the layers might have been hard to understand or derive with data analysis. Here, we will show that this approximation is reasonable due to the smaller numbers of the high-speed trains travelling and their poor effects on the dynamics of the normal-speed one. \newline In our datasets it is possible to identify high-speed trains thanks a specific identifier (``ES*'' for the Italian Network and ``EC'', ``IC'', ``ICE'' for the German Network)  and use them to build the High-speed layer of the Railway Network in a similar way that has been done for the normal layer in the main text. The number of travelling high-speed trains per day is considerably smaller with respect to the normal-speed trains, being of $\sim 210$ and $\sim 1055$ for the Italian and German case respectively. As a consequence also the two networks are smaller compared to the normal-speed ones as shown in table~\ref{tab_1}. The smaller number of nodes indicates the fact that high-speed trains usually connects fewer, more important and distant stations, since it is used mainly for mid-long range movements. This is also reflect in the distribution of the length of the links in the network (Fig.~\ref{fig_distances_dist}), showing a tail which is considerably longer with respect to the normal-speed layer. Other topological properties are similar in the two case, like the degree distribution (Fig.~\ref{fig_deg_dist}) and the associativity coefficient (table~\ref{tab_1}). As an example of the dynamics taking place over the high-speed layer, we show in Fig.~\ref{fig_final_delays} the distribution of the final (positive) delays of high-speed and normal-speed trains. We can see that both for the Italian and German case, the distributions are fat tailed, so that also trains on the high-speed layer can experience large delays and major disruptions.
\begin{table}[h]
	\centering
	\begin{tabular}{||l||l|l|}
		\hline
		& Italian HS Layer &  German HS Layer\\
		\hline
		Number of Nodes & $162$ & $712$  \\
		\hline
		Number of Links & $509$ & $2281$ \\
		\hline
		Avg. Degree & $6.28$ &   $6.41$\\
		\hline
		Degree Assortativity & $0.06$ & $0.15$  \\
		\hline
	\end{tabular}      
	\caption{ Network Metrics of the  High-Speed Layers of the Italian and German Railway Networks.}
	\label{tab_1} 
\end{table}
As a final remark, we validated the approximation of neglecting this layer by looking at the cross-correlations between the time-series of average delays on the links in the networks, similarly to what we have done for the normal-speed layer in the main text. In this case though we checked for correlations not just between the links of the same layer, but also between couples of links from different layers in order to see whether we can spot a signal of a possible inter-layer interaction. Fig.~\ref{fig_crosscorr} shows the cross-correlations in the ``Forward'', ``Backward'' configurations, between the couples of links of the high-speed layer and the couples of links made by a link in the high-speed layer and one in the normal-speed layer. As for the normal-speed layer, decaying correlations exist for non inter-layer couples of links in the Backward configuration, while in all the other cases the signal of correlation is very close to $0$. Hence, it is possible to considered the high-speed and normal-speed layer as independent and non-interacting. It is worth noticing that this measure of correlation might hide possible local interaction effects due to the fact that it is an aggregation of all the couples of links in the network. Such approximation will be then valid when considering global or aggregated metrics (e.g. the delay distributions), but it is not unlikely that more ``fine-grained'' observations (e.g. the distribution of delays on a single link or station) might be influenced by our choice.

\FloatBarrier
\section{Exogenous Delay Distributions}
\begin{figure}[h]
	\centering
	\includegraphics[width=150pt]{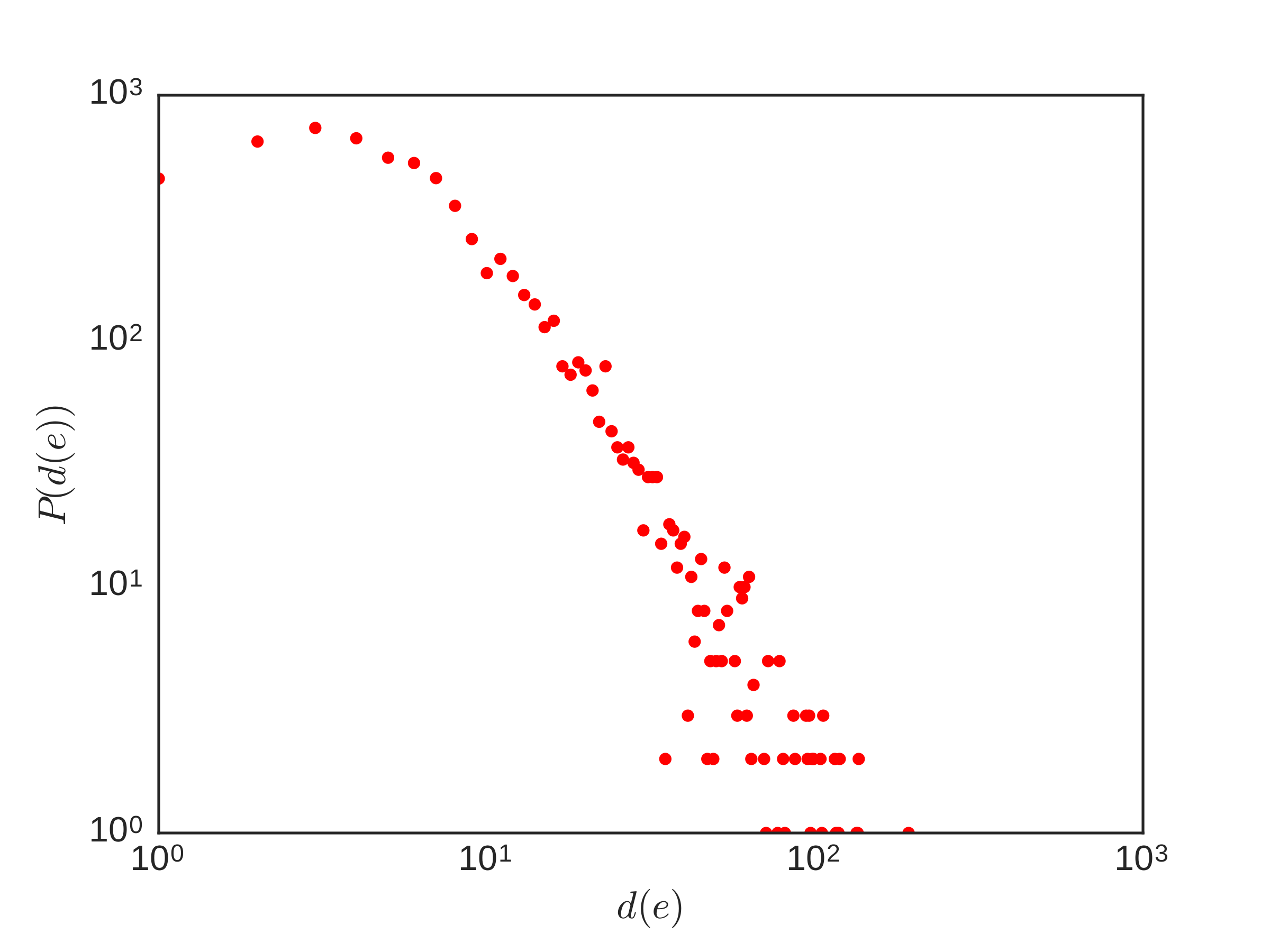}
	\includegraphics[width=150pt]{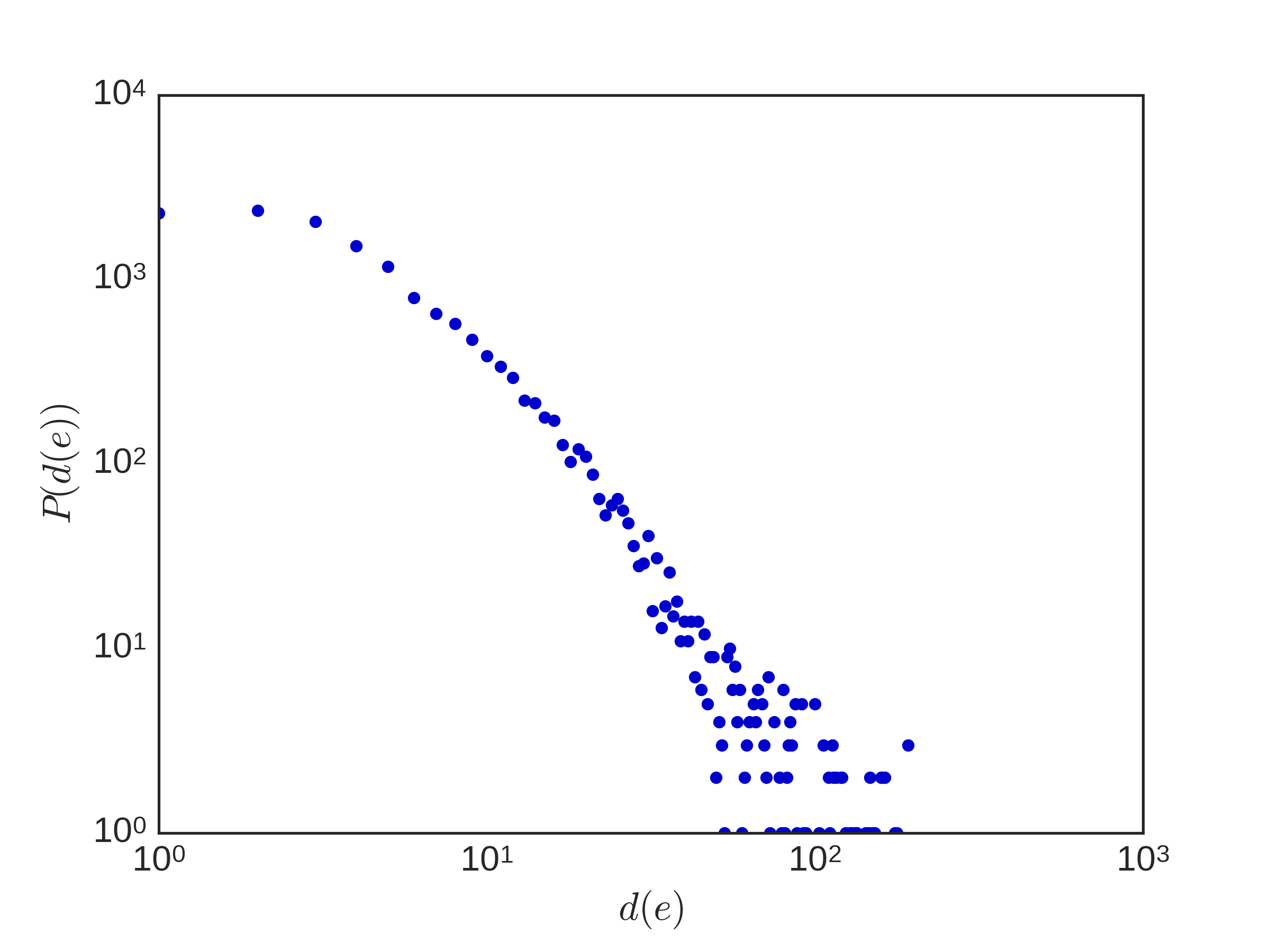}
	\caption{Distributions of the links length $d(e)$ for the Italian (left) and German (right) Railway Networks.}
	\label{fig:links_length}
\end{figure}

\begin{figure}
	\hspace*{-4.3cm}  
	\includegraphics[width=2\textwidth]{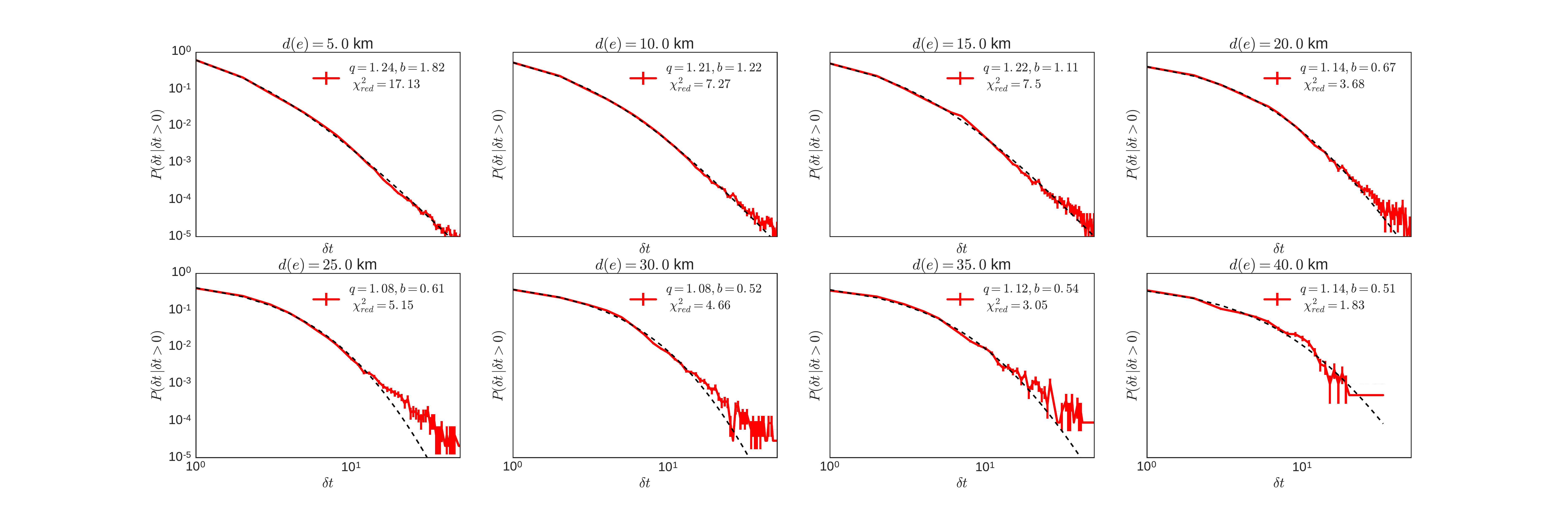}
	\caption{Distributions of the positive exogenous delays according to the length $d(e)$ of the links in the Italian Railway Network. Dotted lines represents the $q$-exponential fit of the distribution. The parameters obtained with the fits are shown in the legend.}
	\label{fig:ITA_exo_pos}
\end{figure}

\begin{figure}
	\hspace*{-4.3cm}  
	\includegraphics[width=2\textwidth]{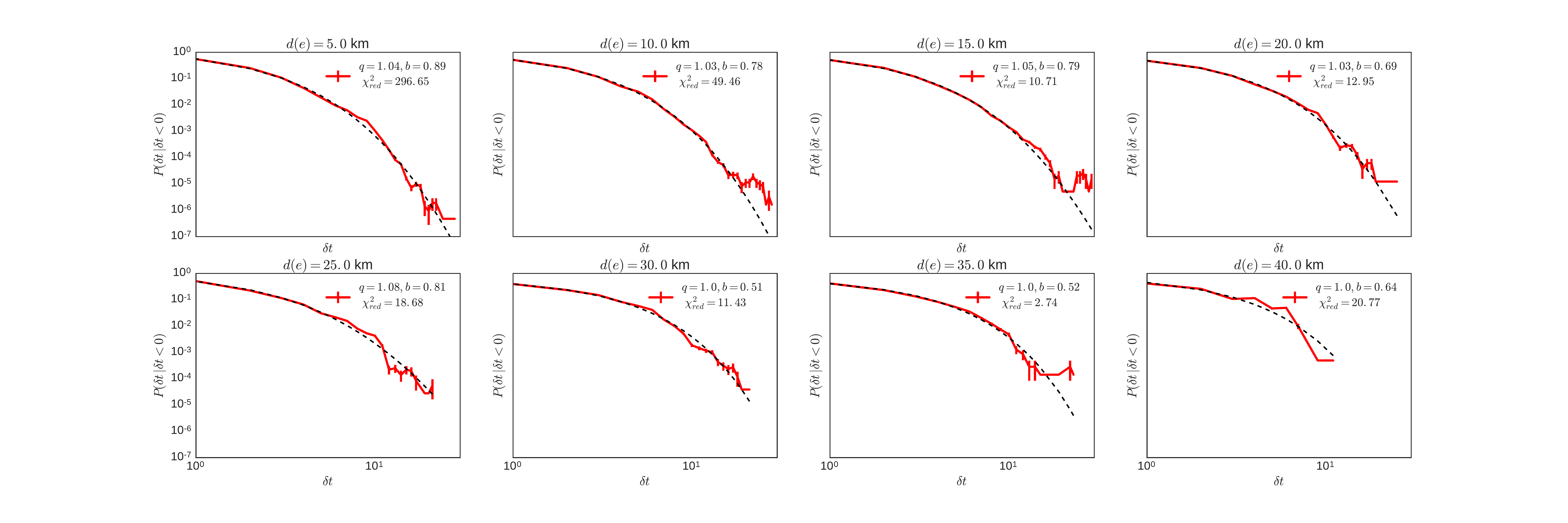}
	\caption{Distributions of the negative exogenous delays according to the length $d(e)$ of the links in the Italian Railway Network. Dotted lines represents the $q$-exponential fit of the distribution. The parameters obtained with the fits are shown in the legend.}
	\label{fig:ITA_exo_neg}
\end{figure}

\begin{figure}
	\hspace*{-4.3cm}  
	\includegraphics[width=2\textwidth]{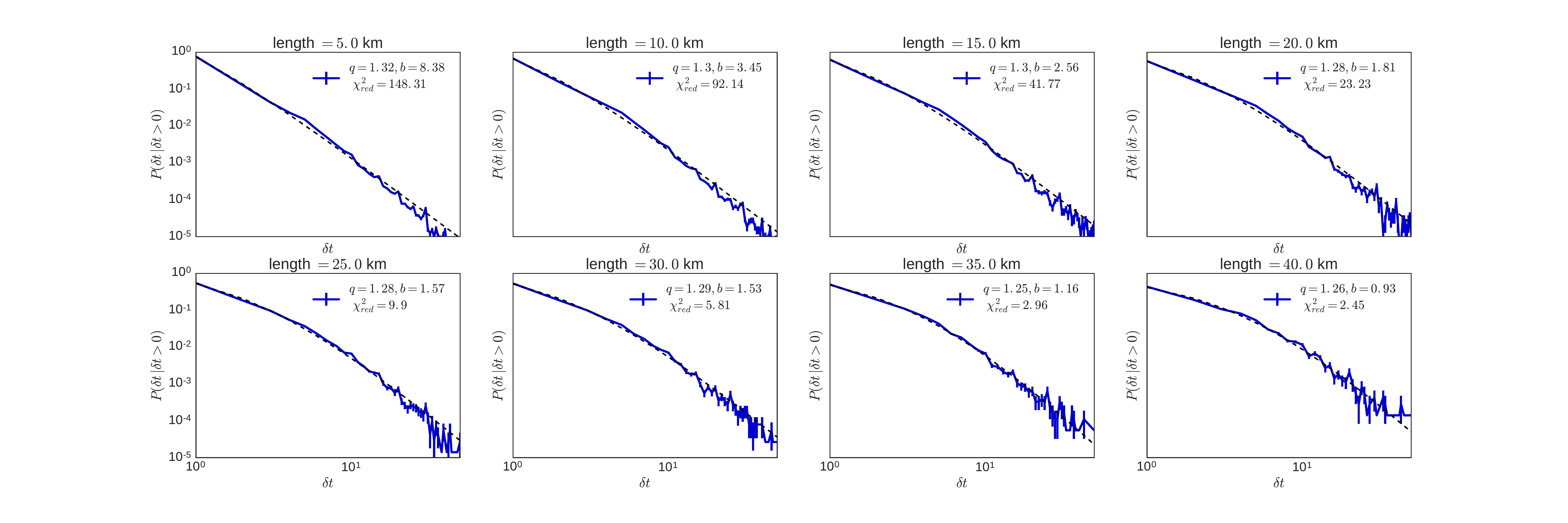}
	\caption{Distributions of the positive exogenous delays according to the length $d(e)$ of the links in the German Railway Network. Dotted lines represents the $q$-exponential fit of the distribution. The parameters obtained with the fits are shown in the legend.}
	\label{fig:GER_exo_pos}
\end{figure}

\begin{figure}
	\includegraphics[width=0.55\textwidth]{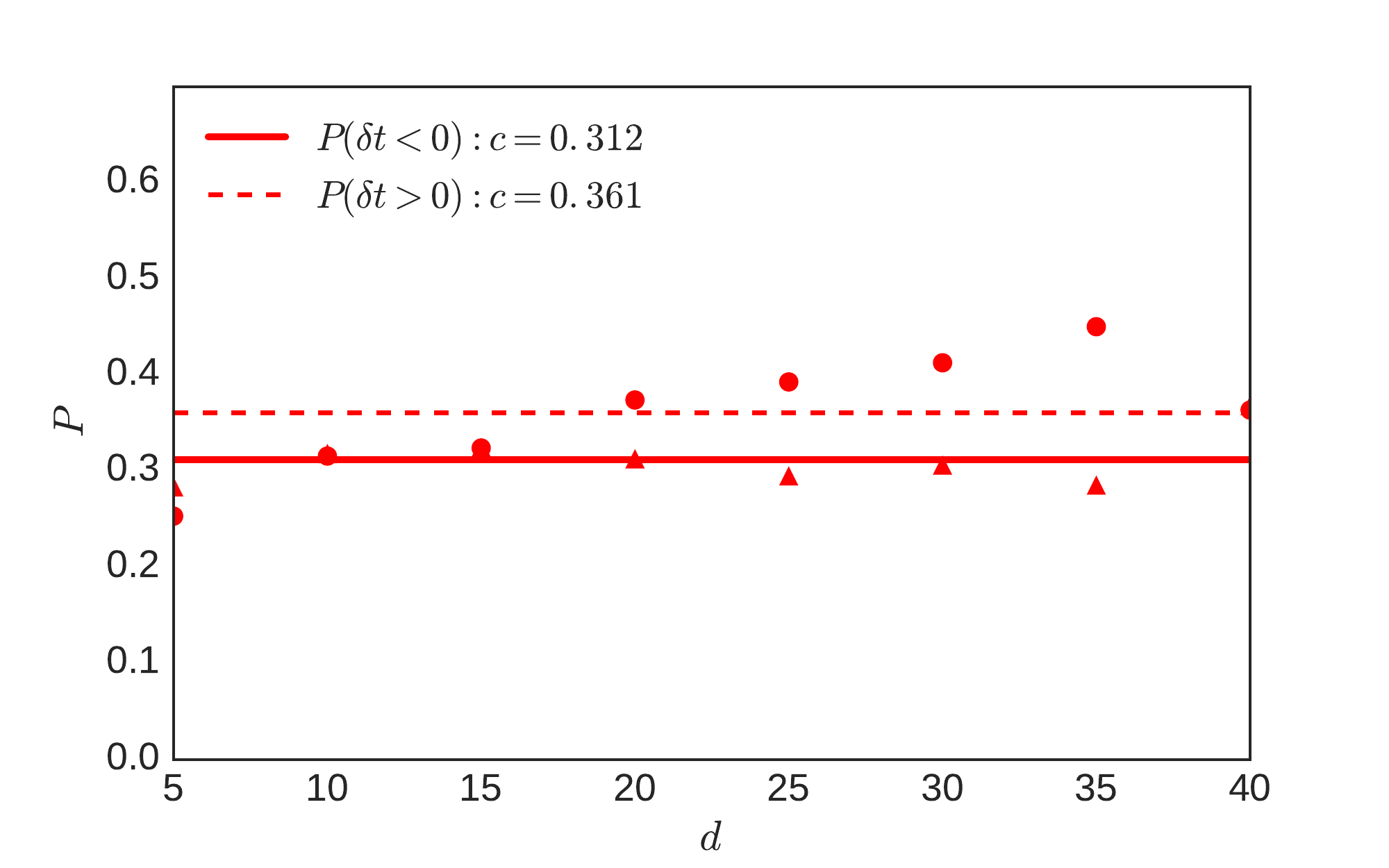}
	\includegraphics[width=0.55\textwidth]{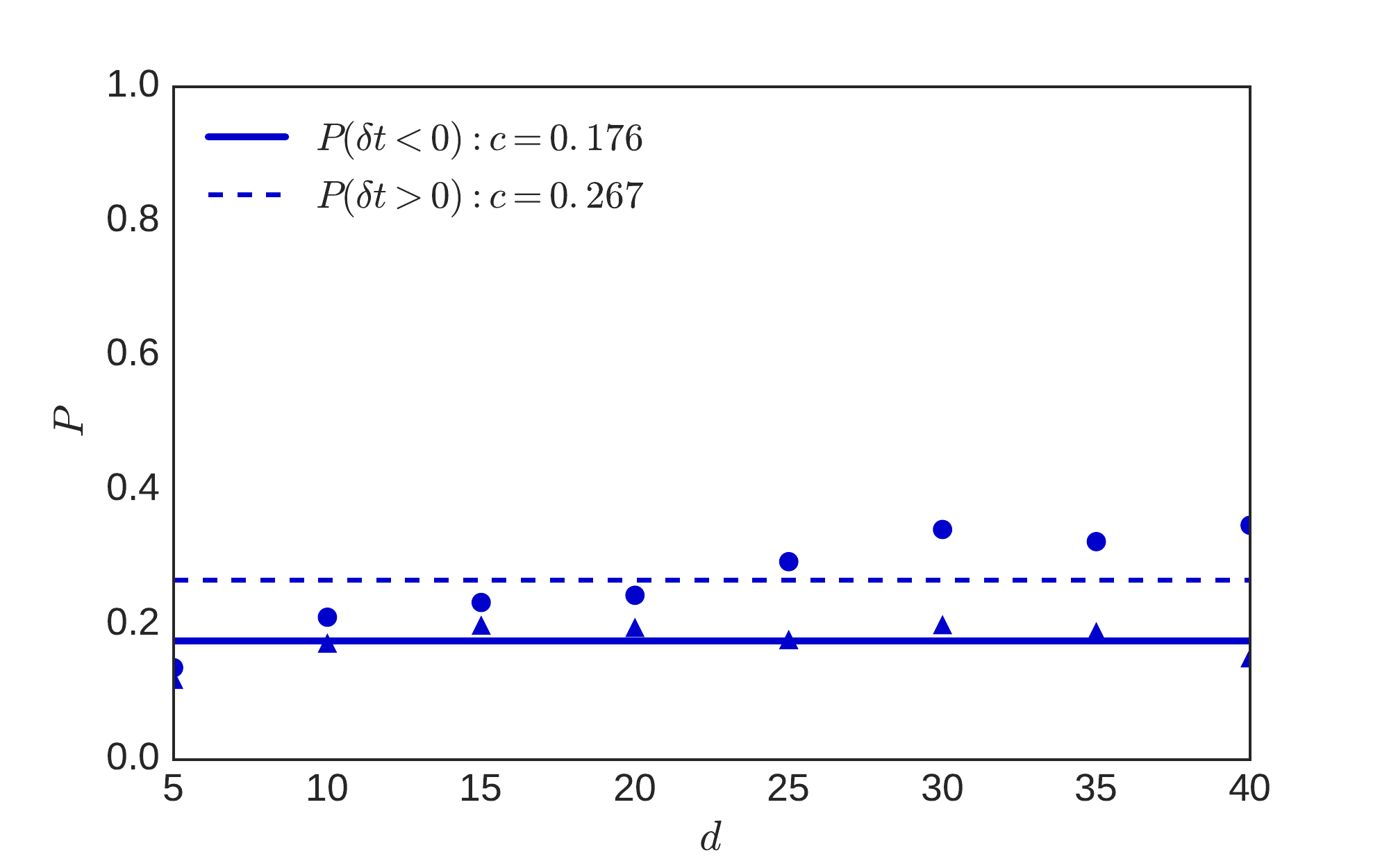}
	\caption{Probability of having a positive (dashed line) and negative (continuous line) exogenous delay as function of the length of the links in the Italian (left) and German (right) Railway Networks.}
	\label{fig:ITA_GER_probs}
\end{figure}

The most trivial way to group the links of the railway networks is according to the geodesic distance between the stations they connect, behind this a rough estimate of the length of the railway between them. Fig.~\ref{fig:links_length}show the distribution of these distances $d(e)$ for all the edges $e$ in the Italian and German Railway Networks. From these distributions we can see that the distances are distributed around a typical value of $\sim 5\,$km, but then span with a long tail until $\sim 100\,$km.
In order to characterize correctly the exogenous delay on the links, we measured the positive and negative exogenous delay distributions aggregating the links according to $d(e)$ as can be seen from Figures~\ref{fig:ITA_exo_pos},~\ref{fig:ITA_exo_neg},~\ref{fig:GER_exo_pos}, and ~\ref{fig:ITA_exo_neg}. In all these cases, we modelled the distribution using a $q$-exponential functional form\cite{Briggs2007modelling,picoli2009q}:
\begin{equation}
e_{q,b}(\delta t) \propto (1 + b(q-1) \delta t)^{1/(1-q)}, q\in[1,2], b>0,
\label{eq:qexp}
\end{equation}
so that in these cases the parameter $q$ and $b$ are depending on $d(e)$.
\begin{figure}
	\hspace*{-4.3cm}  
	\includegraphics[width=2\textwidth]{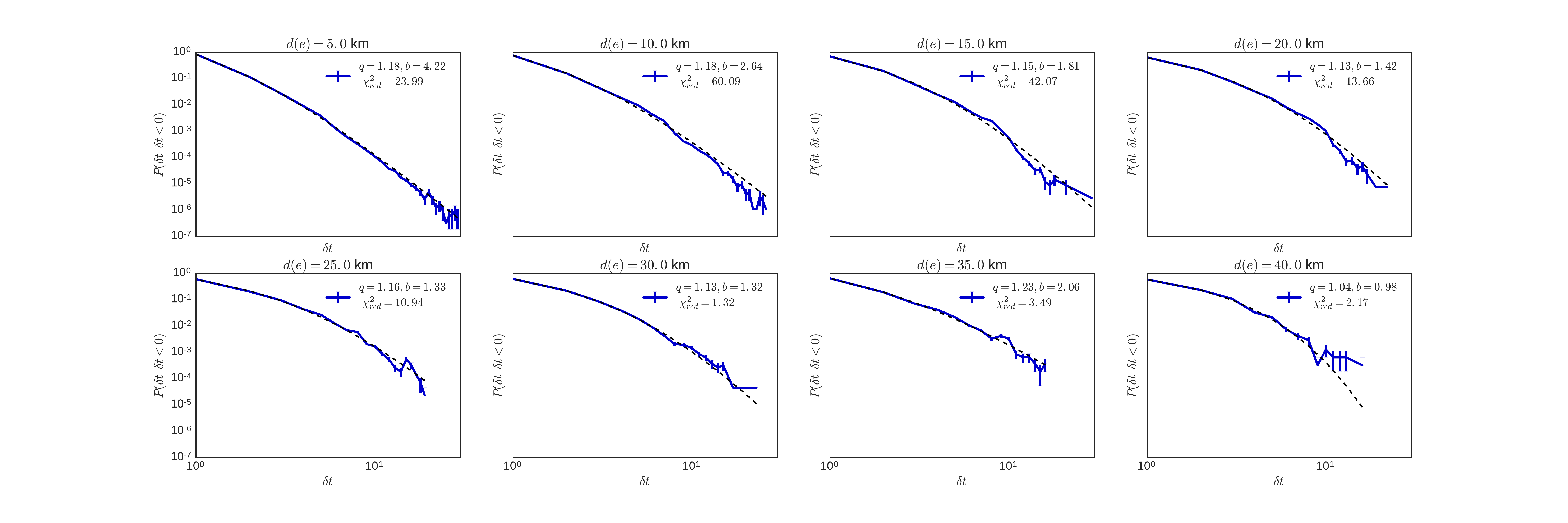}
	\caption{Distributions of the negative exogenous delays according to the length $d(e)$ of the links in the German Railway Network. Dotted lines represents the $q$-exponential fit of the distribution. The parameters obtained with the fits are shown in the legend.}
	\label{fig:GER_exo_neg}
\end{figure}
The behavior of the parameters with respect to $d(e)$ are shown in the main text. We find that in general:
\begin{equation}
q(d) = \textit{const}, \;
b(d) = A d^{-a}.
\label{eq:q_b_links_param}
\end{equation}
The parameters for equation  (\ref{eq:q_b_links_param}) can be found in table~\ref{tab:exo_params}:
\begin{table}[h!]
	\centering
	\begin{tabular}{| l | c | c | c |}
		\hline
		& q & A & a \\ \hline
		ITA positive & $1.15$ & $0.66$ & $5.25$ \\ \hline
		ITA negative & $1.03$ & $0.22$ & $1.33$\\ \hline
		GER positive & $1.28$ & $0.99$ & $37.5$ \\ \hline
		GER negative & $1.15$ & $0.57$ & $0.98$ \\ \hline
	\end{tabular}
	\caption{Parameters for the equation (\ref{eq:q_b_links_param}), governing the behavior of the parameters $q$ and $b$ of the $q$-exponential distribution as the links length $d(e)$ varies.}
	\label{tab:exo_params}
\end{table}
Since these distributions are all conditioned on the fact that the acquired exogenous delay is either positive or negative, we can check whether the probability of these conditions are influenced or not by the length of the link the train is travelling on. Fig.~\ref{fig:ITA_GER_probs} shows these dependencies for both the considered Railway Networks. Despite the fact that a small dependence can be observed in the probability of having positive delays (i.e. it is slightly increasing with $d$), assuming that such probabilities are constant is a good zero-order approximation that we have used in the main text.
\FloatBarrier
\section{Departure Delay Distributions}
\begin{figure}
	\centering
	\includegraphics[width=160pt]{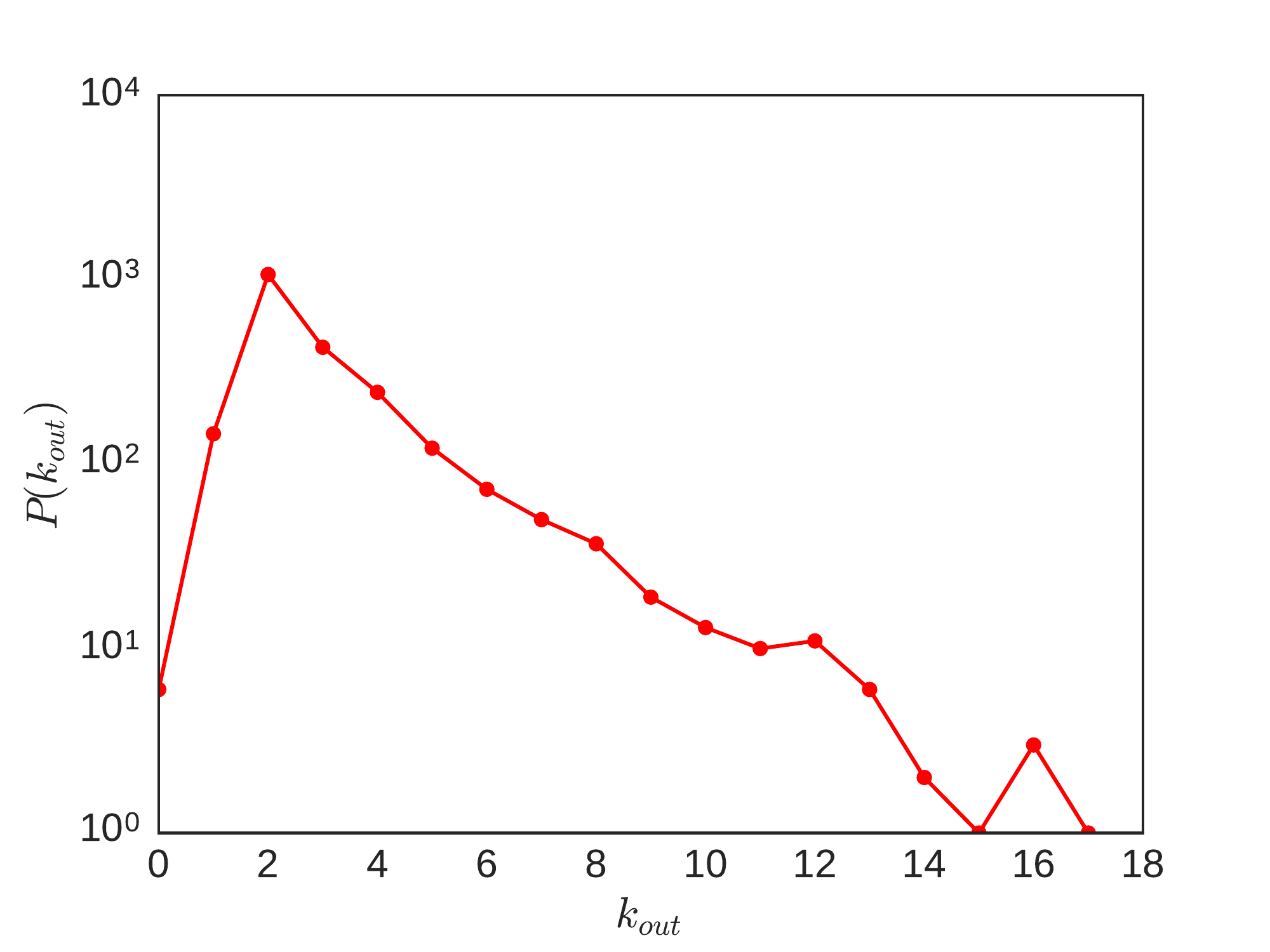}
	\includegraphics[width=160pt]{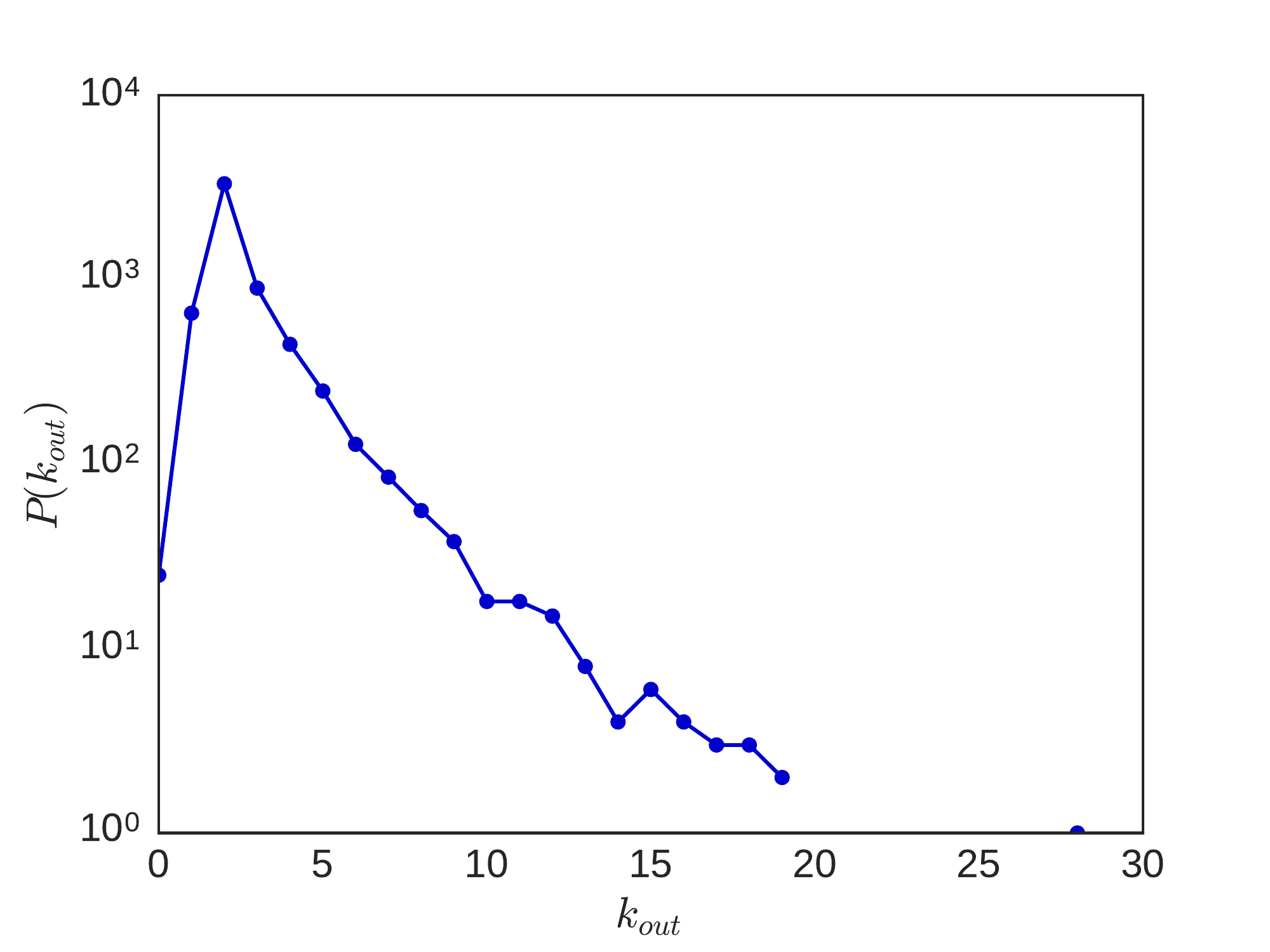}
	\caption{Out-degree distributions for the Italian (left) and German (right) Railway Networks.}
	\label{fig:outdeg_dist}
\end{figure}

\begin{figure}
	\hspace*{-4.3cm}  
	\includegraphics[width=2\textwidth]{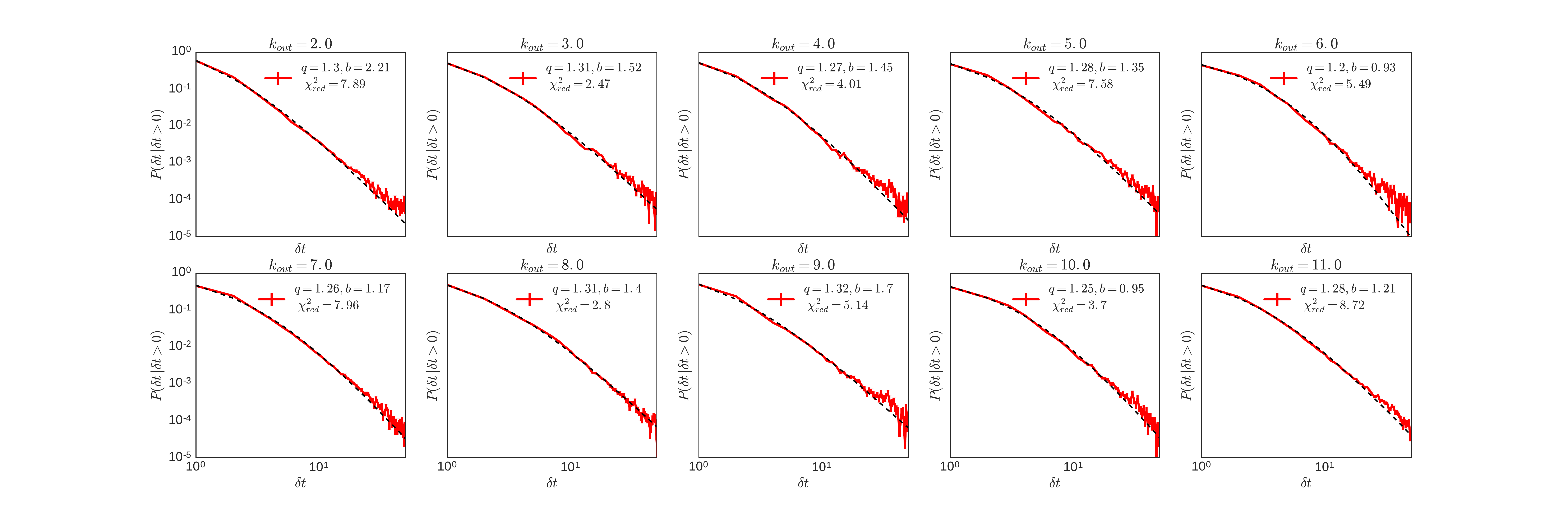}
	\caption{Distributions of the positive departure delays according to $k_{out}$ of the links in the Italian Railway Network. Dotted lines represents the $q$-exponential fit of the distribution. The parameters obtained with the fits are shown in the legend.}
	\label{fig:ITA_dep_pos}
\end{figure}

\begin{figure}
	\hspace*{-4.3cm}  
	\includegraphics[width=2\textwidth]{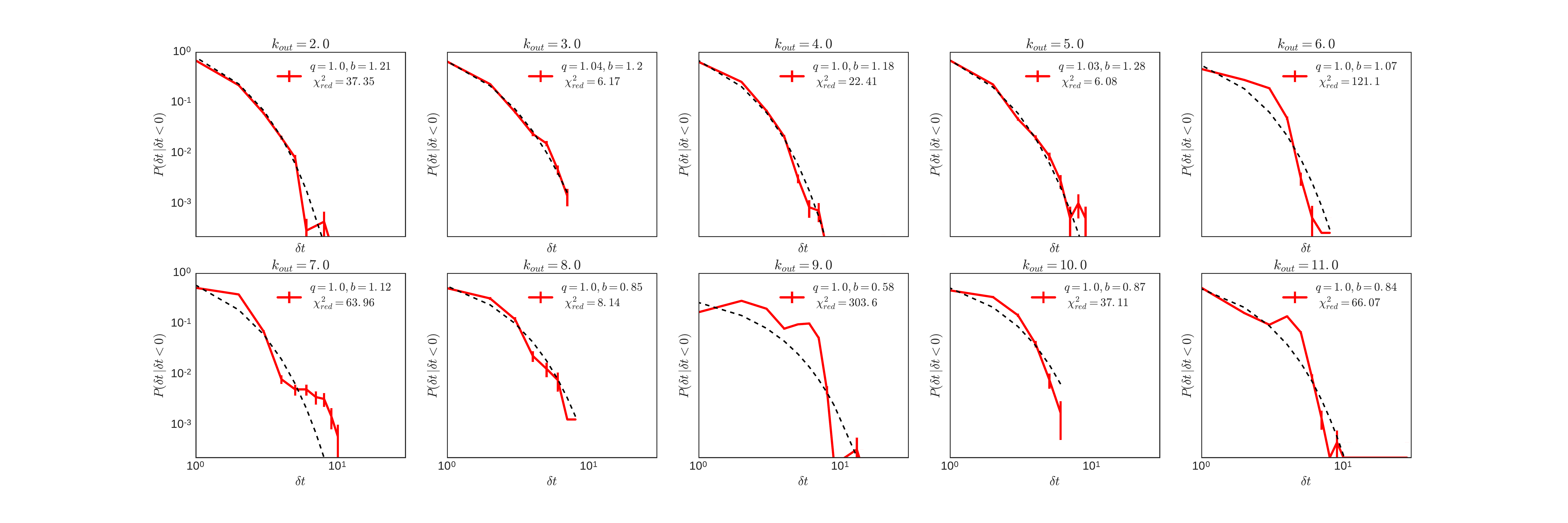}
	\caption{Distributions of the negative departure delays according to $k_{out}$ of the links in the Italian Railway Network. Dotted lines represents the $q$-exponential fit of the distribution. The parameters obtained with the fits are shown in the legend.}
	\label{fig:ITA_dep_neg}
\end{figure}

\begin{figure}
	\hspace*{-4.3cm}  
	\includegraphics[width=2\textwidth]{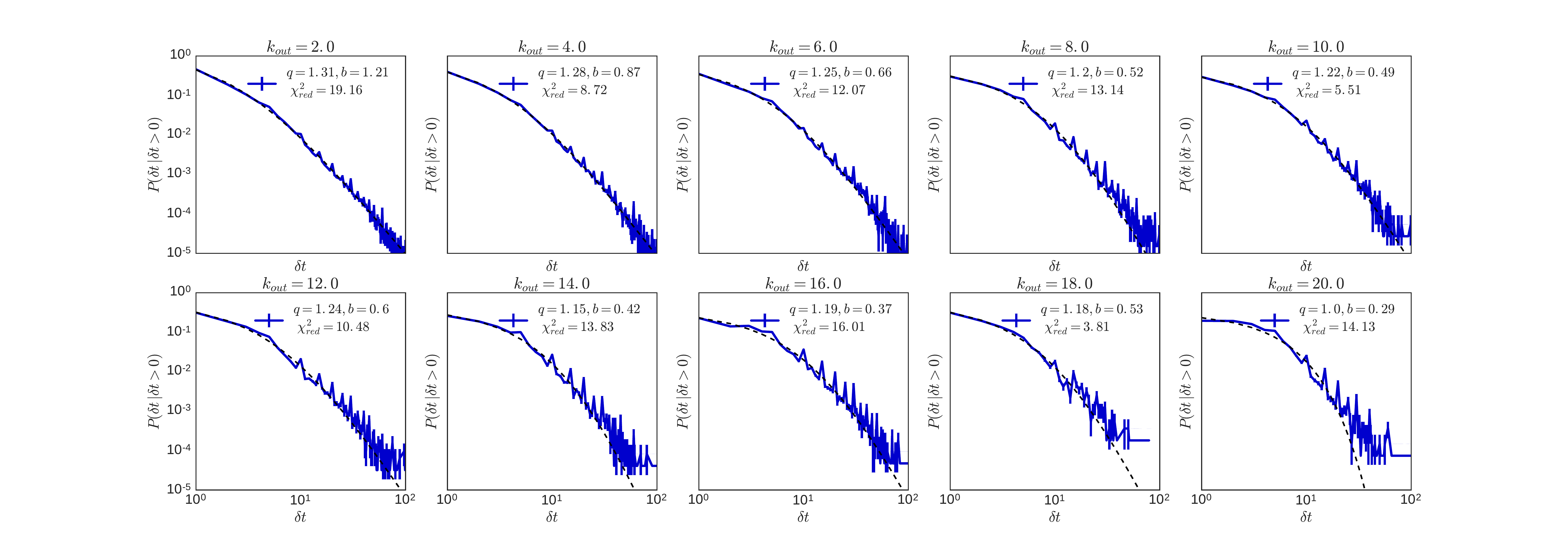}
	\caption{Distributions of the positive departure delays according to $k_{out}$ of the links in the German Railway Network. Dotted lines represents the $q$-exponential fit of the distribution. The parameters obtained with the fits are shown in the legend.}
	\label{fig:GER_dep_pos}
\end{figure}

\begin{figure}
	\includegraphics[width=0.55\textwidth]{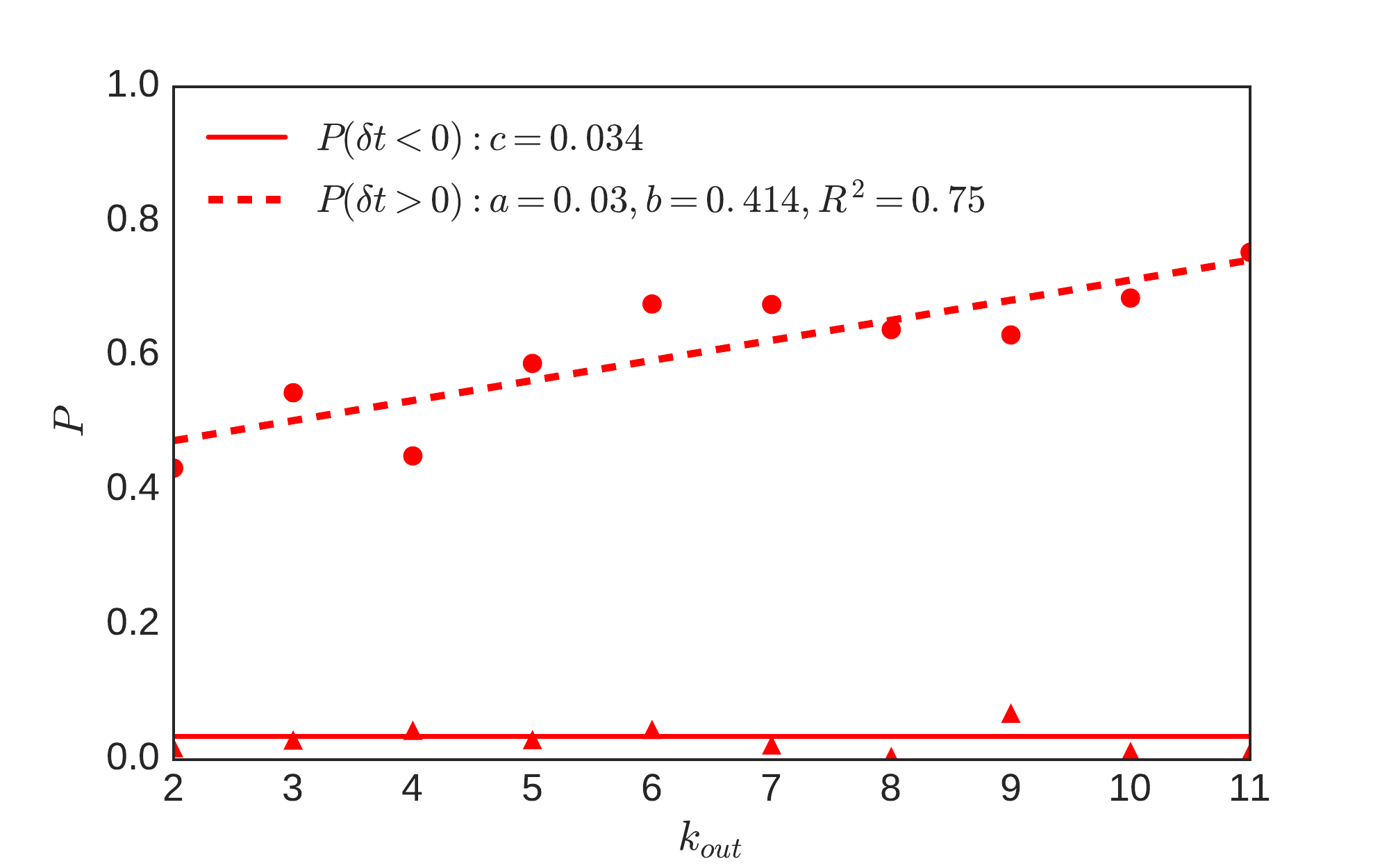}
	\includegraphics[width=0.55\textwidth]{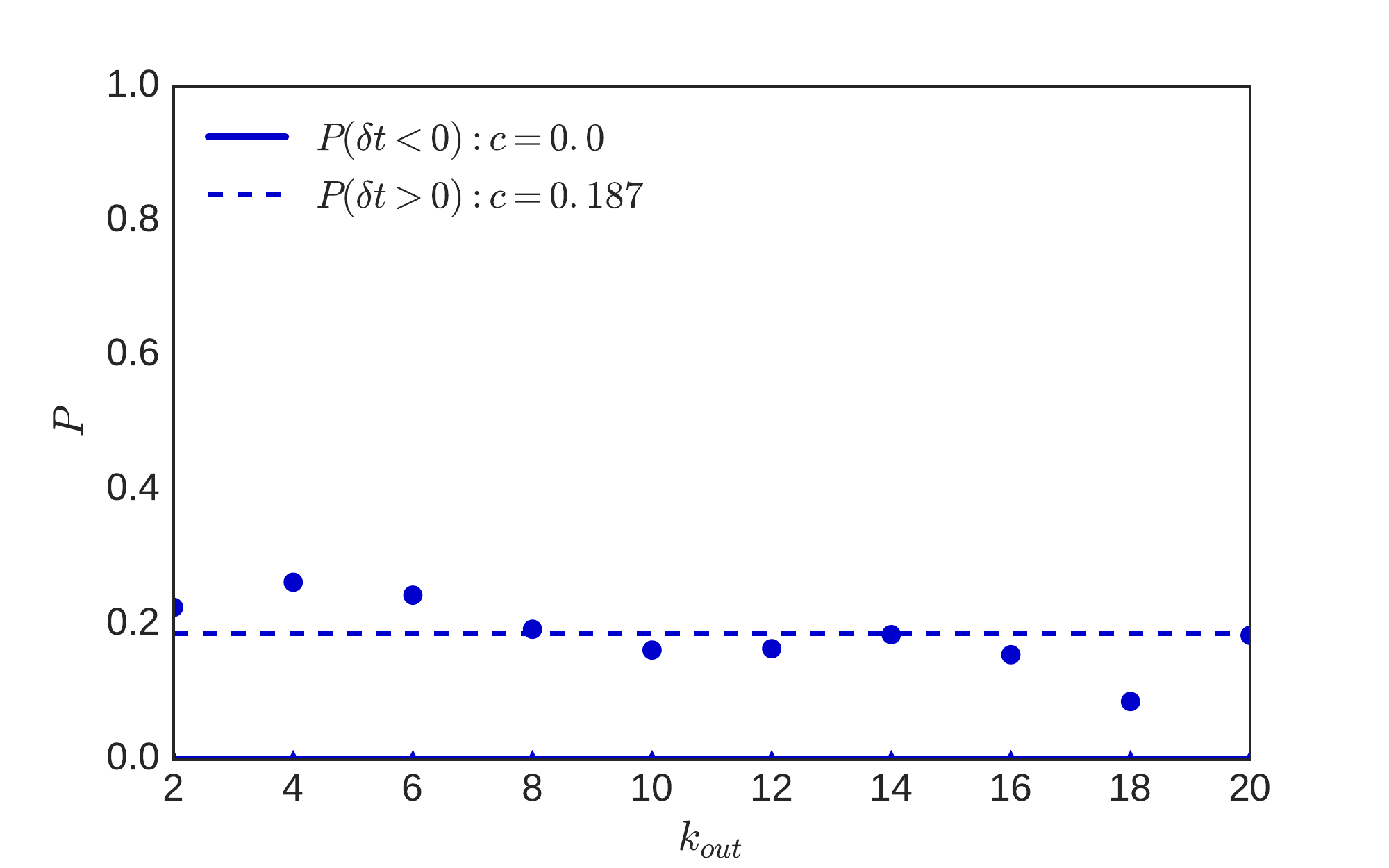}
	\caption{Probability of having a positive (dashed line) and negative (continuous line) departure delay as function of the out-degree of the nodes in the Italian (left) and German (right) Railway Networks. Dashed and continuous lines corresponds to linear or constant fits of the data. We assume that there is no dependency with $k_{out}$ in every case but the positive departure delays in the Italian data where we found the linear relation: $P=ak_{out}+b$ (parameters shown in the legend).}
	\label{fig:ITA_GER_probs_dep}
\end{figure}

An approach similar to the one adopted for the exogenous delays on the links of the networks can also be adopted for the departure delays at the stations. In this case we categorized the departing stations (i.e. a subset of the nodes in the network) according to their out-degree $k_{out}$ whose distributions are shown in Fig.~\ref{fig:outdeg_dist}.
Having divided the nodes of the network according to $k_{out}$, we can fit the aggregated departure delay distributions as $k_{out}$ varies as shown in Fig.~\ref{fig:ITA_dep_neg},~\ref{fig:ITA_dep_pos} and ~\ref{fig:GER_dep_pos}. Note that negative departure delays are not present in the German dataset.
These distribution have been fitted using a $q$-exponential functional form as in equation \ref{eq:qexp}. The behavior of the $q$ and $b$ parameters for these distributions according to $k_{out}$ can be summarized by the equations:
\begin{equation}
q(k_{out}) =\textit{conts},\; b(k_{out}) = Ae^{-a k_{out}}.
\label{eq:q_b_nodes_param}
\end{equation}
The parameters for equations (\ref{eq:q_b_nodes_param}) are obtained by fitting the empirical data as shown in the main text. Table~\ref{tab:dep_params} shows the values obtained with the fit:
\begin{table}[h!]
	\centering
	\begin{tabular}{| l | c | c | c |}
		\hline
		& q & A & a \\ \hline
		ITA positive & $1.28$ & $0.014$ & $1.87$ \\ \hline
		ITA negative & $1.01$ & $0.004$ & $0.99$\\ \hline
		GER positive & $1.20$ & $0.026$ & $1.06$ \\ \hline
	\end{tabular}
	\caption{Parameters for the equation (\ref{eq:q_b_nodes_param}), governing the behavior of the parameters $q$ and $b$ of the $q$-exponential distribution as out-degree $k_{out}$ of the nodes varies.}
	\label{tab:dep_params}
\end{table}
To conclude the investigation about departure delays, it is necessary to study the occurrences of positive and negative ones as $k_{out}$ varies. Fig.~\ref{fig:ITA_GER_probs_dep} shows these dependencies for the Italian and German Railway Networks. Similarly to what we have found for the dependency of the exogenous delay with the length of the links, in the German Railway Network no dependence can be observed and the probabilities of having a positive or negative departure delay can be considered constant in every station. However, this is not true for the Italian Railway Network where just the probability of having a negative delay can be considered constant. On the contrary, the probability of having a positive departure delay increases linearly with $k_{out}$.
\FloatBarrier
\section{Optimal Choice of $\beta$}
\begin{figure}
	\centering
	\includegraphics[width=0.55\textwidth]{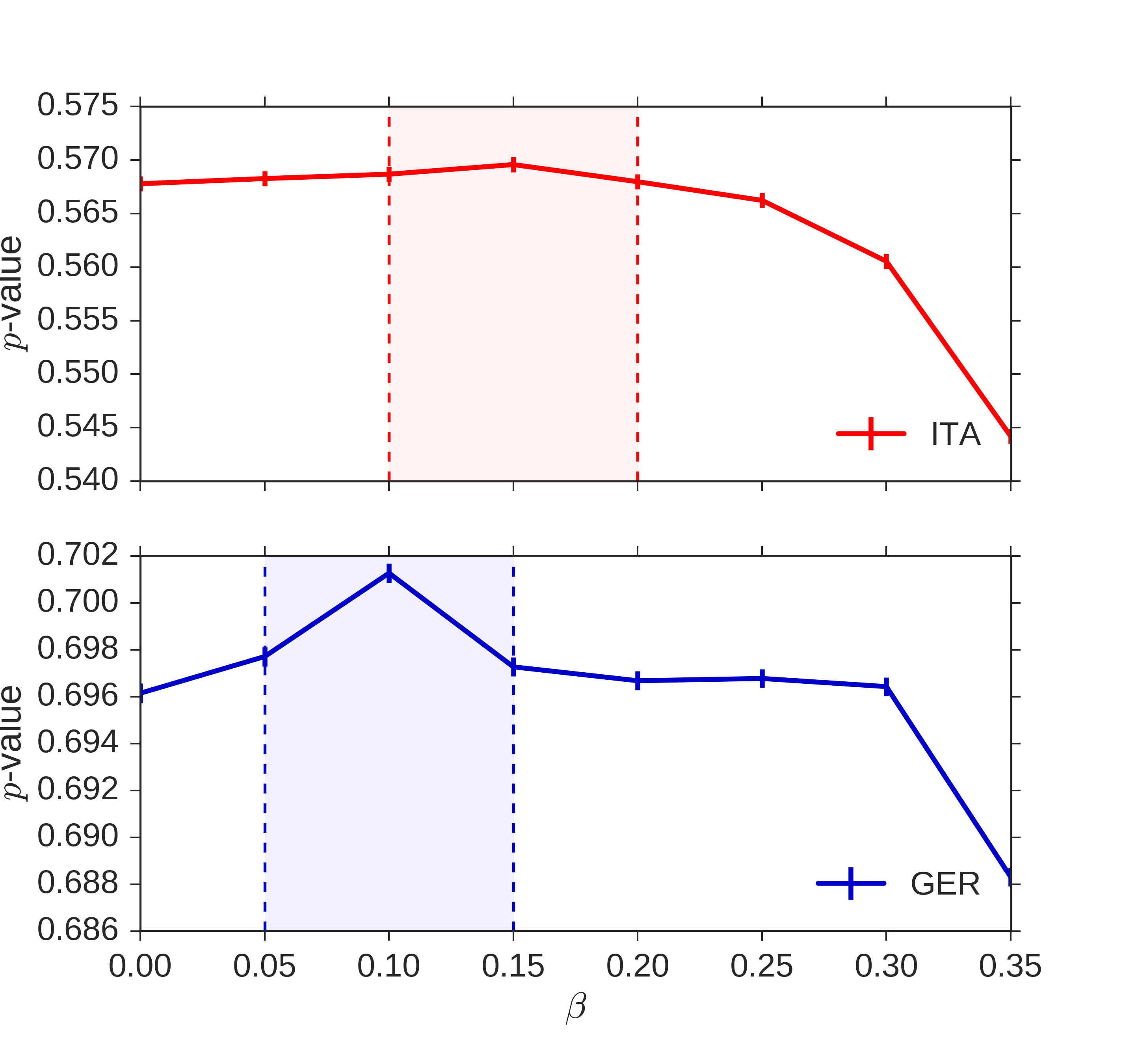}
	\caption{Average $p$-value statistics for the arrival delay for each train at each station in their route as a function of the diffusion parameter $\beta$. The curves have been obtained as the average of the single curve of different days (from the 1$^{\textit{st}}$ to the 6$^{\textit{th}}$ of March 2015). The highlighted regions correspond to the range of values of $\beta$ where the average maximum $p$-value has been observed;}
	\label{fig:optim_p_value}
\end{figure}

In order to determine the optimal value of the $\beta$, we tune our model to reproduce with the highest probability the delay that each train gets whenever it crosses a station during its path. 
Considering a train $i$ arriving at a station $n$ on a given day, we call $\delta t_{i,n}^\mathrm{emp}$ its measured arrival delay at that station as recorded in the dataset. 
Hence, we perform $200$ simulations of the schedule of the considered day in order to compute the distribution $P(\delta t_{i,n})$ of the corresponding $\delta t_{i,n}$. 
Hence, considering the null hypothesis that $\delta t_{i,n}^\mathrm{emp}$ is produced by our model (i.e. it is extracted by $P(\delta t_{i,n})$), we calculate the double tailed $p$-value for such hypothesis for every pair $(i,n)$, i.e. for every train and for every station. For each day we average the $p$-values of all the train-station pairs to obtain the performance metrics for $\beta$.
Fig.~\ref{fig:optim_p_value} shows the dependence of the average $p$-value as a function of $\beta$. The curves have been computed from the simulation of a week of daily schedules. 
The values of $\beta=0.15$ for Italy and  $\beta=0.10$ for Germany allows the model to maximize the probability of reproducing the correct arrival delay for each train at a particular station.
These values will be used in all the numerical simulations in the following, if not otherwise specified.
\FloatBarrier
\section{Predictive limits of the model}
\begin{figure}[htbp]
	\begin{center}
		\includegraphics[width=1.1\textwidth]{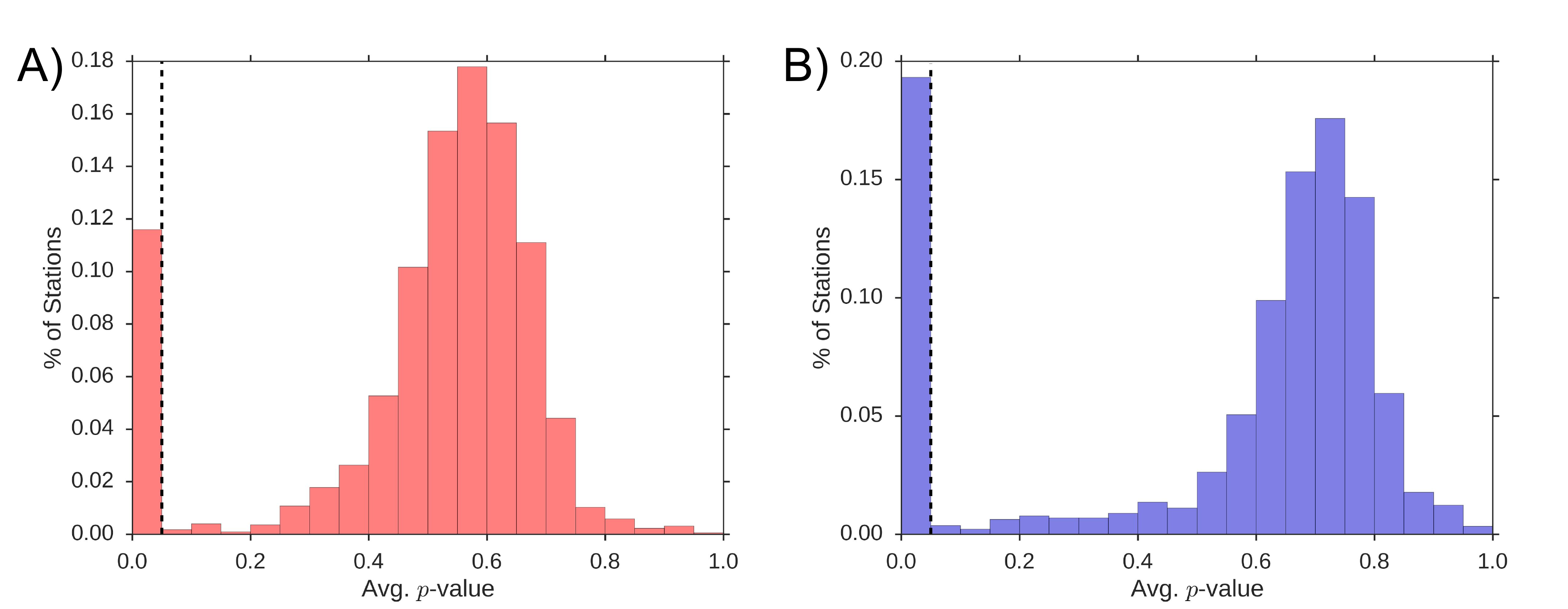}
		\caption{ Average $p$-values distribution for the stations of the Italian (left) and German (right) railway networks with the optimal choice of the diffusion parameter $\beta$. Vertical dashed lines correspond to the value $p=0.05$.
		}
		\label{fig:p_values_distrib}
	\end{center}
\end{figure}

\begin{figure}[htbp]
	\begin{center}
		\includegraphics[width=1.1\textwidth]{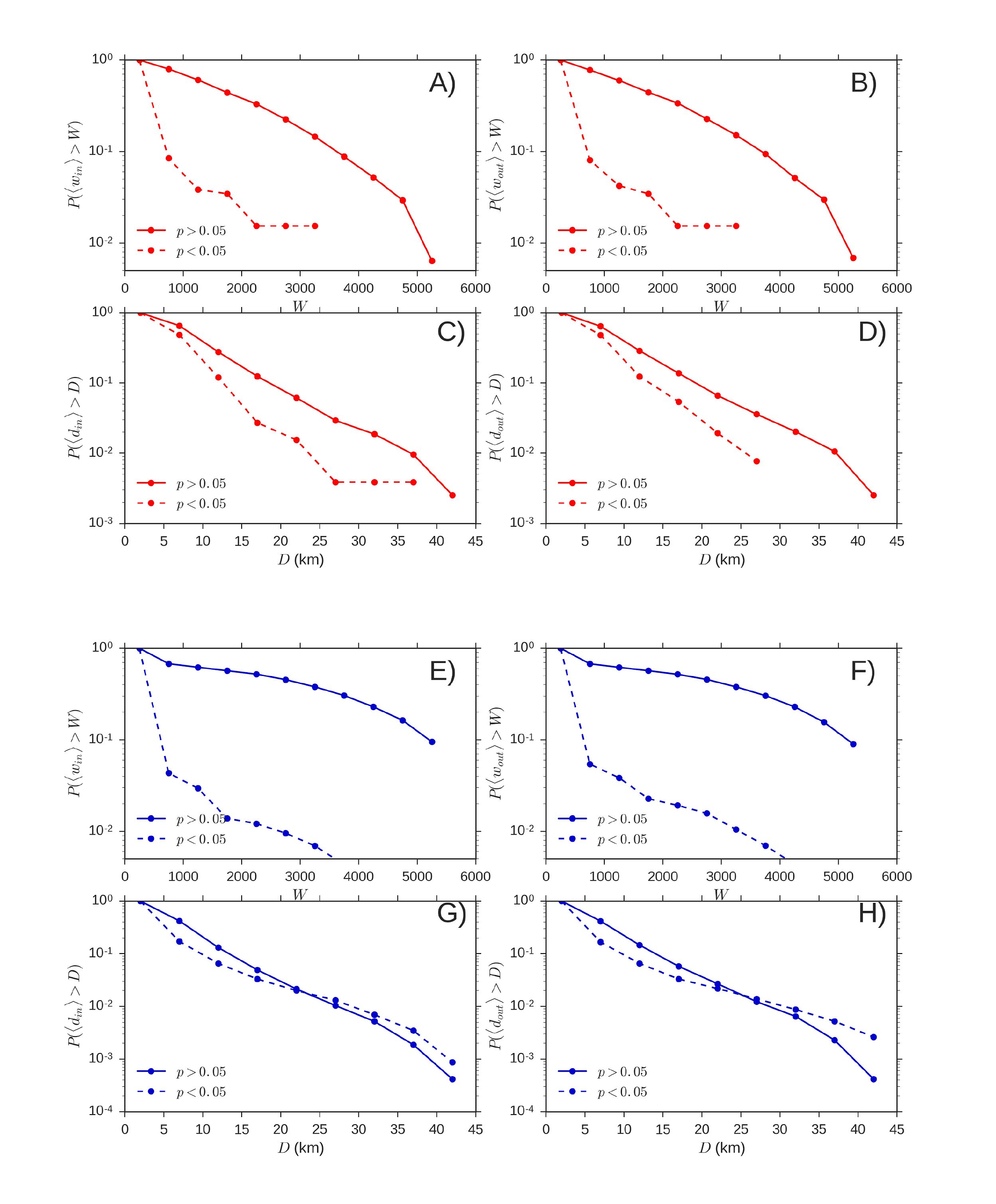}
		\caption{ Distributions of the average weight of the in-going (A-E) and out-going (B-F) of the links of the Railway Networks. Weights are computed as the total number of trains that have travelled over a link during the whole $2015$. Distributions of the average length of the of the in-going (C-G) and out-going (D-H) of the links of the Railway Networks. Red and Blue lines correspond to the Italian and German railway networks, dotted lines are the distributions for the stations with a $p$-value smaller than $0.05$.
		}
		\label{fig:p_values_small}
	\end{center}
\end{figure}

%
By means the definition of the $p$-value of the previous paragraph, we can explore a bit which are the predictive limits of the model, i.e. in which part of the Railway Network it is more likely to not reproduce correctly the delays. 
In other words, we computed for each station $n$ the average $p$-values, by averaging all the $p$-values assigned to the couple $(i,n)$.
Fig.~\ref{fig:p_values_distrib} shows the distribution of these average $p$-values for the stations in the networks, obtained with the optimal value of $\beta$. 
The largest parts of the stations have $p$-values centred around a typical value of $\sim0.6$ for Italy and $\sim 0.7$ for Germany, yet there is a large percentage ($\sim 11\%$ for the Italian and $\sim 20\%$ for the German case) with a $p$-value smaller than $0.05$.
In these stations the predictive power of the model is particularly unsatisfactory and it is interesting to understand something about their features.
\newline 
Fig.~\ref{fig:p_values_small} shows the distribution of the average weight of the links (in terms of the number of trains that have travelled in the links in our whole datasets) and the distribution of the length of the links connected to a station, in the case of stations with $p$-value larger and smaller than $0.05$. 
We can see that in the latter, the distribution of the weight is considerably narrower, indicating that in these stations the traffic is usually low. 
Hence, the disagreement might be the result of a poor sampling of the exogenous disturbances around these stations leading to poor predictions or to a dependence of the transmission parameter $\beta$ on the traffic conditions that have been ruled out when we assumed them to be constant in time and uniform all over the network.
\newline
For the Italian case, these stations also are usually connected to links with a shorter distance. 
This fact can be interpreted as the effect of discrepancies in the fitted exogenous delay model when the links are not sufficiently long.
\FloatBarrier
\bibliographystyle{spphys}       
%

%
\end{document}